\documentstyle[12pt]{article}

\newtheorem{satz}{Satz}[section]
\newtheorem{lemma}[satz]{Lemma}
\newtheorem{definition}{Definition}[section]
\newtheorem{behauptung}[satz]{Behauptung}
\newcommand{\BS}{\begin{satz}}
\newcommand{\ES}{\end{satz}}
\newcommand{\BDEF}{\begin{definition}}
\newcommand{\EDEF}{\end{definition}}
\newcommand{\BB}{\begin{behauptung}}
\newcommand{\EB}{\end{behauptung}}
\newcommand{\BL}{\begin{lemma}}
\newcommand{\EL}{\end{lemma}}
\newcommand{\BD}{\begin{displaymath}}
\newcommand{\ED}{\end{displaymath}}
\newcommand{\BE}{\begin{equation}}
\newcommand{\EE}{\end{equation}}
\newcommand{\BEA}{\begin{eqnarray}}
\newcommand{\EEA}{\end{eqnarray}}
\newcommand{\BEAS}{\begin{eqnarray*}}
\newcommand{\EEAS}{\end{eqnarray*}}
\newcommand{\BA}{\begin{array}}
\newcommand{\EA}{\end{array}}
\newcommand{\w}{\wedge}
\newcommand{\wk}{\stackrel{\times}{\wedge}}
\newcommand{\wm}{\stackrel{\cdot}{\wedge}}
\newcommand{\NN}{\nonumber}
\newcommand{\NI}{\noindent}
\newcommand{\IN}{\indent}
\newcommand{\G}{\left}
\newcommand{\D}{\right}
\newcommand{\GA}{\langle}
\newcommand{\DA}{\rangle}
\newcommand{\DS}{\displaystyle}
\newcommand{\TS}{\textstyle}

\newcommand{\SSS}{\scriptscriptstyle}
\newcommand{\I}{\int_\Sigma}
\newcommand{\IM}{\int_M}
\newcommand{\C}[1]{{\cal #1}}
\newcommand{\V}[1]{{\vec #1}}
\newcommand{\B}[1]{{\bar #1}}
\newcommand{\T}[1]{{\tilde #1}}
\renewcommand{\.}[1]{{\dot #1}}
\renewcommand{\P}{\partial}

\newcommand{\F}{\frac}
\newcommand{\comment}[1]{}
\newcommand{\nummer}[1]{
  \hskip #1
  \refstepcounter{equation}\rlap{\@eqnnum }
  \hskip -#1 }
\newcommand{\setR}{\ifmmode{I\hskip -4pt R}
    \else{\hbox{$I\hskip -4pt R$}}\fi} 
\newcommand{\setQ}{\ifmmode{Q\hskip-5.0pt\vrule height6.0pt depth 0pt
\hskip6pt}
    \else{\hbox{$Q\hskip -5.0pt\vrule height6.0pt depth 0pt\hskip6pt$}}\fi}
\newcommand{\setZ}{\ifmmode{Z\hskip -6pt /}
    \else{\hbox{$Z\hskip -6pt /$}}\fi}
\newcommand{\St}[1]{|_{{}_{{}_{#1}}}}

\renewcommand{\phi}{\varphi}

\newcommand{\R}{\vec{\mbox{Ric}F}}
\newcommand{\ADM}{{\sc adm\ }}

\newcommand{\Tr}{\mbox{tr\hspace{0.5ex}}}

\newcommand{\lra}{\longrightarrow}
\newcommand{\Lra}{\Longrightarrow}
\newcommand{\lmt}{\longmapsto}

\newcommand{\Llra}{\Longleftrightarrow}
\newfont{\euler}{eufm10 scaled \magstephalf}

\newcommand{\RSC}{\setcounter{equation}{0}}
\renewcommand{\S}{\Sigma}

\newcommand{\pr}{\NI{\bf Proof:}}
\newcommand{\ld}{\ldots}
\newcommand{\omegap}{{\B{\omega}^{\SSS +}{}}}

\begin{document}
\begin{center}
{\Large  Triad Formulations of Canonical Gravity without a fixed
reference frame}\\[1cm]
{\large Joachim Schirmer}\\[5mm]
Fakult\"at f\"ur Physik der Universit\"at Freiburg\\
Hermann-Herder-Str. 3, 79104 Freiburg i.Br. / FRG\\
e-mail: schirm@phyq1.physik.uni-freiburg.de\\
FR-THEP 95/8
March 1995
\end{center}

\begin{abstract}
One can simplify the triad formulations of canonical gravity by abandoning
any relation to a fixed coordinate system. That means in case of the \ADM
formalism that one can determine the momentum by direct derivation of the
Lagrange-3-form w.r.t the time-derivative of the triad-1-forms, thus the
momentum is most naturally a 2-form. We apply this concept to the Palatini
formulation where we can closely follow Dirac's concept to find and
eliminate the second class constraints. Following the same way for the
Ashtekar theory it will turn out to be equivalent to two successive
canonical transformations where the first makes explicit use of the spatial
dimension being 3 and the second is usually hidden in the use of
densities. At the end we can give a simple version of the reality
constraints.
\end{abstract}

\section{Technical Preliminaries}
Avoiding any coordinate system will eliminate all determinants from the
theory, but the associated problems will be contained in the frequently
used Hodge operator. Yet the algebra of this operator is simple because our
triads are normalized. To have an effective way of handling this operator
we first introduce some notations and formulas where we mainly follow
the treatment given in \cite{ThII}. Let $(M,g)$ be a $m$-dimensional
pseudo-Riemannian manifold. We first define the interior multiplication $i$
of two forms of different degree.\\
\IN For $q\le p$ let the bilinear mapping $i:\Omega^q(M)\times\Omega^p(M)
\lra \Omega^{p-q}(M); (\mu,\nu)\lmt i_\mu\nu $ have the following properties:
\BEA
   &&i_\mu\nu=g^\sharp(\mu,\nu)=\mu_a\nu_b g^{ab}
             \qquad \mbox{fr } p=q=1\\
  &&i_\mu(\nu_1\w\nu_2)=i_\mu\nu_1\w\nu_2+(-1)^{p_1}\nu_1\w
  i_\mu\nu_2
         \quad \mbox{fr } \nu_i\in\Omega^{p_i}(M),\mu\in\Omega^1(M)\qquad\\
 &&i_{(\mu_1\w\mu_2)}=i_{\mu_2}\circ i_{\mu_1}
\EEA
These properties define $i$ uniquely.
Let $\{a_a\}_{a=1\ldots m}$ be a local basis of the tangential bundle and
$\{a^a\}$ be the dual basis. We will use the following abbreviation for the
basis of $\Omega^p(M)$:
\BE
   a^{a_1\ldots a_p}:= a^{a_1}\w\ldots\w a^{a_p}
\EE
We will not distinguish in the notation between $a_a\in\Gamma_{\mbox{\tiny
loc}}(TM)$ and $a_a^\flat=g_{ab}a^b\in \Omega_{\mbox{\tiny loc}}^1(M)$, so
usually we will omit the $\flat$-sign. Consequently we do not distinguish
between the interior product with a 1-form and the natural injection of the
metric dual vector of this 1-form. This identification is possible unless
variations or derivations get involved. Then one has to keep in mind if for
the definition the metric was used. In a local cobasis the interior product
takes the following form:
\BE
  i_\mu\nu=\F{1}{q!(p-q)!}\mu^{i_1...i_q}\nu_{i_1...i_qj_1...j_{p-q}}
                 a^{j_1...j_{p-q}}
\EE
which is proved in a simple, but tedious calculation. The scalar product of
two $p$-forms is now given by:
\BE
   \GA\mu|\nu\DA:=\,^p\!g^\sharp(\mu,\nu):=i_\mu\nu=i_\nu\mu\qquad
   \mu,\nu\in\Omega^p(M)
\EE
We notice that the interior product $i_\mu$ is the dual mapping of the
exterior multiplication $\mu\w$ with respect to the scalar product $\GA|\DA$:
\BE
   \GA i_\mu\omega|\nu\DA=i_\nu i_\mu\omega=i_{\mu\w\nu}\omega
   =\GA\mu\w\nu|\omega\DA
\EE
We define the Hodge-dual form of a form as interior product with
the canonical volume form of the pseudo-Riemannian manifold
\BE
*:\Omega^p(M)\longrightarrow\Omega^{m-p}(M)\quad\forall p \qquad\quad
*\mu:=i_\mu\eta\qquad,
\EE
so in a local basis we have:
\BE
   *\mu=\F{1}{p!(m-p)!}\mu^{i_1...i_p}
   \eta_{i_1...i_pj_1...j_{m-p}}a^{j_1...j_{m-p}}
\EE
Using the properties of the interior product it is not difficult to prove the
following identities -- here $s$ is the signature of the pseudo-metric:
\BEA
 (i)&&\GA *\mu|*\!\nu\DA=(-1)^s\GA\mu|\nu\DA\\*[0.8ex]
(ii)&&**\!\mu=(-1)^{p(m-p)+s}\mu\qquad\mu\in\Omega^p(M)\\*[0.8ex]
(iii)&&*(\mu\w \nu)=i_\nu*\!\mu\qquad *i_\nu\mu=(-1)^{q(m-q)}*\!\mu\w \nu\quad
        \nu\in\Omega^q(M)\\*[0.8ex]
\label{48}
(iv)&&\mu\w*\nu=\GA \mu|\nu\DA\eta=\nu\w*\mu\quad\nu,\mu\in\Omega^p(M)
\EEA
For an orthonormal co-frame $\{e^i\}_{i=1\ldots m}$ the Hodge operator has
the following useful representation:
\BE
      *e^{i_1...i_p}=\F{1}{(m-p)!}
      \epsilon^{i_1...i_p}{}_{j_{p+1}...j_m} e^{j_{p+1}...j_m}
     = \F{1}{(m-p)!}\eta^{i_1j_1}...\eta^{i_pj_p}\epsilon_{j_1...j_m}
       e^{j_{p+1}...j_m}
\EE
For the exterior derivative of these forms one can prove the formula:
\BE
\label{28}
   d*\!e^{i_1\ldots i_p}=\F{1}{(m-p)!}\epsilon^{i_1\ldots i_p}{}_{j_{p+1}\ld
  j_m}d\,e^{j_{p+1}\ld j_m}= de^l\w *e^{i_1\ld i_p}{}_l
\EE
Let $\omega^a{}_b$ be the connection forms of an arbitrary linear connection
with respect to an arbitrary basis $\{a_a\}$. Then the torsion is given by
\BE
     T^a=Da^a=da^a+\omega^a{}_b\w a^b\qquad,
\EE
so for a torsionfree connection one has
\BEA
   da^a&=&-\omega^a{}_b\w a^b\qquad\mbox{and}\\
   da^{a_1\ld a_p}&=&-\omega^{a_1}{}_b\w a^{ba_2\ld a_p}-\ld
      -\omega^{a_p}{}_b\w a^{a_1\ld a_{p-1}b}
\EEA
The connection forms of a metric linear connection with respect to an
arbitrary basis satisfy
\BE
   g_{ac}\omega^c{}_b+\omega^c{}_a g_{cb}=dg_{ab}\qquad,
\EE
so in an orthonormal frame holds
\BE
     \omega_{ij}+\omega_{ji}=0.
\EE
Using these equations one finds for an arbitrary connection in an orthonormal
basis:
\BEA
\NN  d*\!e^{i_1\ld i_p}&=& de^l\w*e^{i_1\ld i_p}{}_l=
          (T^l-\omega^l{}_k\w e^k)\w*e^{i_1\ld i_p}{}_l\\
\label{27}
         &=& T^l\w*e^{i_1\ld i_p}{}_l
             +\omega_l{}^{i_1}\w*e^{li_2\ld i_p}+\ld
            +\omega_l{}^{i_p}\w*e^{i_1\ld i_{p-1}l}
\EEA
and hence for an arbitrary basis:
\BEA
\NN d*\!a^{a_1\ld a_p}&=&T^b\w*a^{a_1\ld a_p}{}_b+(\omega^{ba_1}+dg^{ba_1})\w
     *a_b{}^{i_2\ld i_p}+\ld\\
    &&+(\omega^{ba_p}+dg^{ba_p})\w *a^{a_1\ld a_{p-1}}{}_b-(\omega^a{}_a
    -{\TS\F{1}{2}}g^{ab}dg_{ab})\w*a^{a_1\ld a_p}
\EEA
For the special case of the Levi-Civit…-connection this formula reads:
\BE
\label{26}
  d*\!a^{a_1\ld a_p}= -\omega^{a_1}{}_b\w*a^{ba_2\ld a_p}-\ld
                    -\omega^{a_p}{}_b\w*a^{a_1\ld a_{p-1}b}
\EE
For the convenience of the reader we give the expression of Christoffel
symbols of the Levi-Civit…-connection in an arbitrary basis:
\BEA
 \omega_{a\,bc}:=g_{ad}\omega^d{}_c(a_b)
            &=&{\TS\F{1}{2}}\Big(g_{ab,c}-g_{bc,a}+g_{ca,b}
                    +C_{a\,bc}-C_{b\,ca}+C_{c\,ab}\Big)      \\
\NN  C_{a\,{bc}}&=&g_{ad}a^d([a_b,a_c])=-g_{ad}da^d(a_b,a_c)
\EEA
It is possible to represent the connection form of the Levi-Civit…
connection
by the inner product of an orthonormal basis and its derivatives
\BEA
\NN    \omega_{ij}&=&{\TS\F{1}{2}}\G(i_jde_i-i_ide_j-i_{ij}de_k\cdot e^k\D)\\
\NN       &=&i_jde_i-i_ide_j-{\TS\F{1}{2}}i_{ij}(de_k\w e^k)\\
       &=&(-1)^{m+s}*\G[-e_j\w*de_i+e_i\w*de_j+{\TS\F{1}{2}}e_{ij}\w*(de^k\w
       e_k)\D]
\EEA
In the following paragraphs one important point is the functional derivative
with respect to forms. We will illustrate this method with a well known
example. Instead of varying the metric components with respect to a fixed
coordinate basis we will vary the orthonormal frame, i.e. we will vary the
four 1-forms which are declared to be orthonormal. So varying the
metric can be represented by varying with respect to forms. Let us consider
the Einstein-Hilbert-action, as usual the variation shall vanish on the
boundary of $M$:
\BEA
   S[e^\mu]&=&\F{1}{2}\int_M R_{\mu\nu}\w*e^{\mu\nu}\\
\NN \delta S&=&\F{1}{2}\int_M\G[\delta\!R_{\mu\nu}\w*e^{\mu\nu}
               +R_{\mu\nu}\w \delta *\!e^{\mu\nu}\D]\\
\NN        &=&\F{1}{2}\int_{\P M}\delta\omega_{\mu\nu}\w *e^{\mu\nu}
                   +\int_M R_{\mu\nu}\w \delta e^\rho\w *e^{\mu\nu}{}_\rho
\EEA
\vspace{-8pt}
\BEA
\NN   \F{\delta S}{\delta e^\rho}
         &=& \F{1}{2}R_{\mu\nu}\w*e^{\mu\nu}{}_\rho
           =\F{1}{2}\bigg[
           \GA R_{\mu\nu}|e^{\mu\nu}\DA*\!e_\rho+
            \GA R_{\mu\nu}|e^\nu{}_\rho\DA*\!e^\mu+
            \GA R_{\mu\nu}|e_\rho{}^\mu\DA*\!e^\nu\bigg]\\
\label{9}
           &=&-*\G(R_\rho-{\F{1}{2}}Re_\rho\D)
\EEA
Here
\vspace{-12pt}
\BE
   R_\nu:=i^\mu R_{\mu\nu}
\EE
is the Ricci-form, which is symmetric, i.e. $\GA R_\mu|e_\nu\DA =\GA
R_\nu|e_\mu\DA$ and
\BE
\label{4}
   R:=i^\mu R_\mu=i^\mu i^\nu R_{\mu\nu}
\EE
is the Ricci-scalar.It will be useful in the last paragraph to use instead
of the orthonormal triads the dual frame of the $m\quad (m-1)-$forms
$*e^\mu$. We will show that one can functionally derive the expression
$\int_M e^i\w \alpha_i$, with respect to $*e^i$ and obtain a well-defined
1-form. Let $\{\alpha_i\}_{i=1\ld m}$ be $m~~(m-1)-$forms independent of
$e$ and define the $m$ 1-forms $\{\beta_i\}_{i=1\ld m}$ by:
\BE
     *e^j{}_i\w\beta_j=\alpha_i
\EE
Then one has
\vspace{-12pt}
\BEAS
  &\DS \IM e^i\w\alpha_i
    =\IM e^i\w*e^j{}_i\w\beta_j=(m-1)\IM*e^j\w\beta_j&\\[16pt]
  &\delta\alpha_i=\delta*\! e^j{}_i\w\beta_j+*e^j{}_i\w\delta\beta_j=0&
\EEAS
\BEAS
 \DS\delta\IM e^i\w\alpha_i
       &=&(m-1)\IM\delta*\!e^j\w\beta_j+\IM e^i\w*e^j{}_i\w\delta\beta_j\\
 \DS &=&(m-1)\IM\delta\!*\!e^j\w\beta_j-\IM e^i\w\delta*\!e^j{}_i\w\beta_j\\
 \DS    &=&(m-1)\IM\delta*\!e^j\w\beta_j
             -\IM e^i\w\delta e^k\w *e^j{}_{ik}\w\beta_j\\
 \DS       &=&(m-1)\IM\delta*\!e^j\w\beta_j-(m-2)\IM\delta
           e^k\w*e^j{}_k\w\beta_j
\EEAS
\vspace{-12pt}
\BEAS
 \DS  \delta\IM e^i\w\alpha_i&=&\IM\delta\!*\!e^j\w\beta_j\\
 \DS  \F{\delta}{\delta*\!e^j}\IM e^i\w\alpha_i&=&\beta_j
\EEAS
The equation $\qquad   *e^j{}_i\w\beta_j=\alpha_i\qquad $ is easily solved:
\BEAS
 &e_k\w*e^j{}_i\w\beta_j=-*\!e_i\w\beta_k+\eta_{ki}*\!e^j\w\beta_j
       =e_k\w\alpha_i&\\
 &-\GA \beta_k|e_i\DA+\eta_{ki}\GA\beta_j|e^j\DA=(-1)^s\GA e_k|*\!\alpha_i
    \DA&\\
 &\DS\GA\beta_j|e^j\DA=\F{(-1)^s}{m-1}\GA e_j|*\!\alpha^j\DA&\\
 &\DS\GA\beta_i|e_j\DA=(-1)^s\G[\F{\eta_{ij}}{m-1}\GA e_k|*\!\alpha^k\DA
      -\GA e_i|*\!\alpha_j\DA\D]&\\
 &\DS\beta_i=(-1)^s\G[\F{e_i}{m-1}\GA e_k|*\!\alpha^k\DA
     -\GA e_i|*\!\alpha_j\DA e^j\D]&
\EEAS
Thus we have proved the formula
\BE
  \F{\delta}{\delta *\! e^i}\IM e^k\w\alpha_k
  =(-1)^s\G[\F{e_i}{m-1}\GA e_k|*\!\alpha^k\DA-\GA e_i|*\! \alpha_j\DA
  e^j\D]
  \qquad,
\EE
which we specialize for the case of $\dim M=3,\ s=0$:
\BE
\label{16}
   \F{\delta}{\delta *\!e^i}\IM e^k\w\alpha_k=-\epsilon_i{}^{lm}i_l\alpha_m
      -\F{1}{2}e_i\GA\alpha_m|*\!e^m\DA
\EE

\section{The ADM formulation in triads}\RSC
The space-time manifold is assumed to be a parametric set of imbedded
spacelike hypersurfaces $\Sigma_t$, where we call the non-unique
parametrization "time"; i.e. there exists a diffeomorphism $i:\setR\times
\Sigma\lra M$, which can be used for identification. Let $(x^1,x^2,x^3)$ be a
coordinate system for $\Sigma$, then $(t,x^1,x^2,x^3)$ is a chart for $
\setR\times\Sigma$ and $(\B t,\B x^1,\B x^2,\B x^3):=(t,x^1,x^2,x^3)\circ
i^{-1}$ is one for $M$. We will generally distinguish the analogous
quantities on $M$ and $\S$ by using a bar for quantities defined on $M$.
So if not defined in another way one obtains the unbarred quantities by
pullback
with the map $i_t(x):=i(t,x),~i_t:\Sigma\lra M.$ A pseudo-metric $\B g$ --
signature (-1,+1,+1,+1) -- on $M$ defines a normal $\B e_0$  to the
subspace of the tangential space which is spanned by  $\{\F{\P}{\P\B
{x^i}}\}_{i=1,2,3}$, it is the normal to the $\{\B t=const\}$-surfaces
$\S_t=i_t\circ\S$ in $M$. With respect to this normal we decompose the time
vector field $\F{\P}{\P\B t}$:
\vspace{-10pt}
\BE
     \F{\P}{\P \B t}=\B N\B e_0+\B{\V N}
\EE
At this point it seems that the use of a certain identification $i$ leads
to a gauge fixing of lapse and shift, because it determines at once the
time vectorfield and the normal. But we use this concept only to find the
3+1-decomposed Lagrangian which corresponds to the covariant action
principle. When we have obtained it, we will turn our point of view and try
to reconstruct the space-time metric from the decomposed data. If we do not
know the space-time-metric, the normal on the right-hand-side of the
equation will be unknown. By varying lapse and shift, we will vary the
normal, and thus it is no surprise that the constraints associated with
lapse and shift are equivalent to Einstein's equations restricted to the
normal, i.e. $i_{\B e_0}\B G=0$. Now let $\{\B e_\mu\}_{\mu=0\ld 3}$ be
adapted orthonormal tetrades on $M$, such that $\{\B e_i\}_{i=1,2,3}$ are
always parallel to the hypersurfaces $\Sigma_t$. The metric $\B g$ will be
represented by the dual cotetrades $\{\B e^\mu\}_{\mu=0\ld 3},~\B
g=\eta_{\mu\nu}\B e^\mu\otimes \B e^\nu$. Using the imbedding $i_t$ we can
pull back the tetrades $e^\mu=i_t^*\B e^\mu, ~i_t^*\B e^0=Ni_t^*d\B t=0$ We
first decompose the Einstein-Hilbert Lagrange form\footnote{Greek letters
are summed from 0 to 3, Latin letters from 1 to 3}:
\BEA
\NN \B {\C L}&=&\frac{1}{2}\G(\B R_{\mu\nu}\wedge*\B e^{\mu\nu}\D)=
                 \frac{1}{2} \G(2\B R_{0i}\wedge*\B e^{0i}
                 +\B R_{ij}\wedge*\B e^{ij}\D)\\
\NN  &=&\frac{1}{2} \left[2\,d\B \omega_{0i}\!\wedge*\B e^{0i}+
        2\,\B \omega_{0k}\!\wedge\B \omega^k_{~i}\wedge*\B e^{0i}+\,
        ^3\!\B R_{ij}\wedge\!*\B e^{ij}
        +\B \omega_{i0}\wedge\B \omega^0_{~j}\!\wedge\!*\B e^{ij}\right]\\
\NN  &\stackrel{(\ref{26})}{=}&
         \frac{1}{2}\left[2\,d\!\left(\B \omega_{0i}\wedge*\B e^{0i}\right)-
        2\B \omega_{0i}\wedge\B \omega^0_{~j}\wedge*\B e^{ji}+\,
        ^3\!\B R_{ij}\wedge*\B e^{ij}+
        \B \omega_{0i}\wedge\B \omega_{0j}\wedge*\B e^{ij}\right]\\
     &=&\frac{1}{2}\left[2\,d\!\left(\B \omega_{0i}\wedge*\B
     e^{0i}\right)+\,
                (^3\!\B R_{ij}-\B \omega_{0i}\wedge\B \omega_{0j})\wedge*\B
                e^{ij}\right]
\EEA
Neglecting the exact form which turns into a surface integral after
integration, we can write the action in the 3+1-decomposed form:
\BEA
\NN  S(\B e)=\int_M \B{\C L}
     &=&\frac{1}{2}\int_M \G(\,^3\!\B R_{ij}-\B\omega_{0i}\w\B\omega_{0j}
     \D)\w*\B e^{ij}\\
   &=&\frac{1}{2}\int dt\I i_t^*i_{\partial/\partial \bar{t}}
      \G(\,^3\!\B R_{ij}-\B \omega_{0i}\w\B \omega_{0j}\D)\w*\B e^{ij}
\EEA
Here $i_{\P/\P\B t}$ is the natural injection of a vector field in a form.
Using $\B e^0=\B N\,d\B t$ and $i_t^*d\B t=0$, this reduces to
\vspace{-8pt}
\BEA
\NN S(\B e)&=&\F{1}{2}\int_\setR dt\I i_t^*(^3\!\B R_{ij}-\B\omega_{0i}
           \w\B\omega_{0j})
     \w i_t^*i_{\P/\P\B t}\stackrel{4}{*}\B
e^{ij}\\
      &=&\F{1}{2}\int_\setR dt\I N(R_{ij}-\omega_{0i}\w\omega_{0j})
       \w\stackrel{3}{*}e^{ij}\qquad,
\EEA
where we have identified ${i_t^*}^3\!\B R_{ij}$ and $R_{ij}$, which is
defined on  $\S$ as Levi-Civit… curvature form to the metric defined by the
triads $\{e^i\}$. Using the relation $\omega_{0i}=-i_iK$, where $K$  is the
$(0,2)$-tensor of the extrinsic curvature, one can easily recognize the
form given above as the \ADM action-integral:
\vspace{-8pt}
\BEA
\NN  S(e,\. e)&=&\F{1}{2}\int dt\I N
       \GA R_{ij}-\omega_{0i}\w\omega_{0j}|e^{ij}\DA\eta\\
    &=&\F{1}{2}\int dt\I N(R-K^i{}_iK^j{}_j+K^i{}_jK^j{}_i)\eta
\EEA
We now define the Lagrange function as follows -- the index 3 on the
Hodge-operator will be understood for the rest of the paragraph.
\BE
      L(e^i,\dot{e}^i,N,\V N)=\I \F{N}{2}(R_{ij}-\omega_{0i}\w\omega_{0j})\w*
      e^{ij}
\EE
To make the dependance of $\dot{e}^i$ obvious it is necessary to use the
relation between the time-derivative of the triads and the extrinsic
curvature:
\BE
   \dot{e}^i=\F{d}{dt}\St{s=t}(i_s^*\B e^i)=\F{d}{ds}\St{s=0}
    \G(i_t^*{\Phi_s^{\P/\P\B t}}^*\B e^i\D)=i_t^*L_{\P/\P\B t}\B e^i
\EE
because it holds $i_{t+s}=\Phi_s^{\P/\P\B t}\circ i_t$, where
$\Phi_s^{\P/\P\B t}$ denotes the flow of the time-vectorfield
$\F{\P}{\P\B t}$. So we take the Lie-derivative and pull back the result:
\BEA
\NN L_{\P/\P\B t}\B e^i&=&L_{\B{\V N}} \B e^i+\B Ni_{\B e_0}d\B e^i\\
\NN  &=&L_{\B{\V N}}\B e^i-\B N\B \omega^i{}_\mu(\B e_0)\B e^\mu
     +\B N\B\omega^i{}_0\\
\label{1}
    \dot{e}^i=i_t^*L_{\P/\P\B t}\B e^i&=&L_\V Ne^i-a^i{}_je^j
    +N\omega^i{}_0\\
^\label{13}
    a^i{}_j&:=&Ni_t^*\B \omega^i{}_j(\B e_0)
\EEA
Here $a^i{}_j$ is the rotational parameter which is characteristic for a
triad theory. This relation corresponds to the equation
\BE
\label{2}
  \dot{q}_{ab}=L_\V N q_{ab}+2NK_{ab}
\EE
in the usual \ADM theory, which could be derived from (\ref{1}). One is
tempted to treat the rotational parameter $a$ like shift and lapse as an
additional variable, which turns out to be a gauge parameter, since its time
derivative does not appear in the Lagrangian. But this will lead to
redundancies: If in (\ref{2}) $\dot{q},~N$ and $\V N$ are given, one can
determine $K_{ab}$, and if in (\ref{1}) $\dot{e},~N$ and $\V N$ are given,
one can determine $\omega$ and $a$. So one can not arbitrarily choose $a$ to
determine $\omega_{0i}$ from $\dot{e}$. The reason is the symmetry of the
extrinsic curvature, which is a consequence of the fact that our original
connection on $M$ was torsionfree.
\BEA
\NN   \B T^0=0& \Lra &i_t^*\B T^0=i_t^*(d\B e^0+\B \omega^0{}_\mu\w\B e^\mu)
      =\omega^0{}_i\w e^i=0\\
\label{6}
      &\Lra& \GA\omega_{0i}|e_j\DA=\GA\omega_{0j}|e_i\DA
\EEA
We define the symmetric and antisymmetric part of the 1-forms $\dot{e}^i$
and $L_\V N e^i$ in the following way
\BEA
    {\dot{e}^i}_{S/A}&:=&{\TS\F{1}{2}}(\dot{e}^i\pm\GA\dot{e}^j|e^i\DA e_j)\\
    L_\V Ne^i_{S/A}&:=&{\TS\F{1}{2}}(L_\V Ne^i\pm\GA L_\V Ne^j|e^i\DA e_j)
\EEA
and can split the equation (\ref{1}) in these parts:
\BEA
\label{33}
   \dot{e}^i_S={L_\V Ne^i}_S-N\omega_0{}^i\\
\label{47}
   \dot{e}^i_A={L_\V Ne^i}_A-a^i{}_je^j
\EEA
This split will not be possible in the Palatini-case, where the connection
is not assumed to be torsionfree and thus the extrinsic curvature is not
symmetric. In the Palatini-case the rotational parameter $a$ will become a
variable comparable with lapse $N$ and shift $\V N$.

We notice that only $\dot{e}^i_S$ appears in the Lagrange function. We can
now either introduce a fixed basis $\{a_a\}_{a=1\ld 3}$ and find a variable,
such that $\dot{e}^i_S$ is a proper time derivative -- $\dot{e}^i_S$ is not
the time derivative of $e^i_S=e^i$. Then we end up with the metric \ADM
formulation
\vspace{-10pt}
\BEA
    q_{ab}&=&\eta_{ij}e^i(a_a)e^j(a_b)\\
 \dot{q}_{ab}&=&2\GA \dot{e}_i{}_S|e_j\DA e^i(a_a)e^j(a_b)\quad\footnotemark
\EEA
\footnotetext{Here $\dot{e}_j$ denotes $\eta_{ij}\dot{e}^i$, not
$(\dot{e}_i)^\flat$, they differ by a sign.}
and of course we can write the Lagrangian in terms of $q_{ab}$ and
$\dot{q}_{ab}$. But we can also disregard the fact, that only the symmetric
part of $\dot{e}^i$ appears in the Lagrangian and expect a primary
constraint. The momentum form is easily found:
\BE
\label{35}
    p_i:=\F{\delta L}{\delta\dot{e}^i}=\omega_{0j}\w*e_i{}^j
        =\GA\omega_{0j}|e^j\DA*e_i-\GA\omega_{0j}|e_i\DA*\!e^j
\EE
The momentum is naturally a vector-valued 2-form, generally in a space of
dimension $n$ it is a $n-1$-form. This is analogous to the usual \ADM
formulation, where the momentum-form is a (0,2)-tensorvalued 3-form, where
one usually splits the 3-form into a density and $d^3 x$. We define the
contracted momentum
\BE
   p:=\GA p_i|*\!e^i\DA=2\GA\omega_{0i}|e^i\DA
\EE
and would like to reexpress the extrinsic curvature $\omega_{0i}$ by a
simple calculation as follows
\BE
\label{7}
   \omega_{0i}=-\GA p_j|*\!e_i\DA e^j+{\TS\F{1}{2}}pe_i,
\EE
but this would be inconsistent, if $\GA p_i|*\!e_j\DA\not=\GA p_j|*\!e_i\DA$,
since we know that the extrinsic curvature is symmetric. Thus it is only
possible to obtain $\omega_{0i}$ from $p_i$,  if $p_i\w
e_j=p_j\w e_i$.This is a consequence of the nonsolvability of the equation
\BE
\label{3}
   p_i-\F{\delta L}{\delta\dot{e}^i}(e,\dot{e}^i)
     =p_i+\F{1}{N}\G(\dot{e}^k{}_S-L_\V Ne^k{}_S\D)\w *e_{ik}=0\qquad,
\EE
which has to be regarded as a constraint equation. Multiplying this
constraint by $e_j$ and taking the antisymmetric part yields the necessary
consistency condition
\BE
     p_i\w e_j-p_j\w e_i\approx0
\EE
which is an equivalent formulation of the constraint implied by the
solvability of (\ref{3}). So this is a primary constraint implied by the
symmetry of the extrinsic curvature, which is a consequence of the fact
that the fourdimensional connection to be constructed from $e^i$ and
$\dot{e}^i$ is torsionfree.We can now perform the Legendre-transform.
\BEA
\NN  H&=&\I p_i\w\dot{e^i}-L\\
\NN   &=&\I p_i\w{\dot{e^i}}_S+\I p_i\w{\dot{e}^i}_A
        -\I{\TS\F{N}{2}}(R_{ij}-\omega_{0i}\w\omega_{0j})\w*e^{ij}\\
\label{20}
      &=&\I p_i\w L_\V Ne^i- \I a^i{}_j p_i\w e^j+
          \I{\TS\F{N}{2}}\G[\GA p_i|*\!e^j\DA\GA p_j|*\!e^i\DA
            -{\TS\F{1}{2}}p^2-R\D]\eta
\EEA
The significance of $a$ is here at first that of a Lagrange multiplier of
the rotational constraint. But if one want to reconstruct the metric
space-time one has to relate $a$ to a space-time quantity as given in
equation (\ref{13}), in order that the Hamiltonian equation for $\. e$ and
the geometric equation (\ref{1}) agree. It seems that the Hamiltonian
description does not determine the quantity $b_i:=Ni_t^*\B\omega_{0i}(\B
e_0)$, but we will show in the following paragraph, that one has
$b_i=-\GA dN|e_i\DA$(\ref{19}). Since the time dervatives of lapse $N$ and
shift $\V N$ do not appear in the Lagrangian one finds the following
primary constraints:
\BE
    p_N=\F{\delta L}{\delta\dot{N}}\approx 0\qquad\qquad
      p_{\dot{\V N}}=\F{\delta L}{\delta \.{\V N}}\approx 0\qquad,
\EE
which give rise to the following secondary constraints:
\BE
\label{30}
 \BA{rcccl}
    C_H&=&\{H,p_N\}&=&\eta\G[\GA p_i|*\!e^j\DA\GA p_j|*\!e^i\DA
       -{\TS\F{1}{2}}p^2-R\D]\approx0 \\
  {C_D}_i&=&\{H,p_{N^i}\}&=&-dp_i+p_j\w i_ide^j
      =-Dp_i-\omega^j{}_k(e_i)p_j\w e^k\approx 0
  \EA
\EE
where in the case of the diffeomorphism constraint a surface term had
to be  neglected. Using the rotational primary constraint
\BE
   {C_R}^j{}_i={\TS\F{1}{2}}(p^j\w e_i - p_i\w e^j)  \qquad.
\EE
we could redefine the diffeomorphism constraint in a concise way
$Dp_i\approx 0$, but this will not be possible in the Palatini case, and
since we prefer to work with integrated constraints where our
diffeomorphismconstraint has the simple meaning of the momentum mapping of
the action of the diffeomorphism group of $\S$, we stick to the original
version. But this is a matter of taste and we will have the choice also
in the Ashtekar formulation. We can represent the Hamiltonian as sum of the
integrated constraints
\BE
  \BA{rcccccl}
 \hspace{-0.2cm}
  H(e,p,N,\V N,a)&=&\I N^i{C_D}_i(e,p)+\int_{\P\S}N^j p_j
       &+&\I a^i{}_j{C_R}^j{}_i&+&\I NC_H\\
  &:=&H_D(e,p,\V N)&+&H_R(e,p,a)&+&H_H(e,p,N)
  \EA
\EE
and we will call the integrated versions of the constraints like the
constraints itself, diffeomorphism, rotational and Hamiltonian constraint.
We can now determine the equations of motion:
\BEA
\label{32}
   \dot{e}^i=\F{\delta H}{\delta p_i}&=&L_\V N e^i-a^i{}_je^j
       +N(\GA p_j|*\!e^i\DA e^j-{\TS\F{1}{2}}pe^i)  \\
\NN  \dot{p}_i=-\F{\delta H}{\delta e^i}&=&L_\V N p_i+a^j{}_ip_j
      -N*\!\G(R_i-{\TS\F{1}{2}}Re_i\D)+*(\nabla\!_idN-e_i\Delta N)\\
\label{24}
    &&-N\G(\GA p_i|*\!e^j\DA p_j-{\TS\F{1}{2}}pp_i\D)
    +\F{N}{2}\G(\GA p_j|*\!e^k\DA
    \GA p_k|*\!e^j\DA-{\TS\F{1}{2}}p^2\D)*\!e_i\qquad
\EEA
The Ricci-form $R_i$ is defined as in (\ref{4}) and $\Delta N$ denotes as
usual the Laplacian of the lapse function. In the equation for the momentum
surface terms had to be neglected in order to derive the first and the last
term. We will show the contribution of the Hamiltonian part for the equation
of motion.
\BEAS
  &\delta\G[\GA p_i|*\!e^j\DA\GA p_j|*\!e^j\DA\eta\D]
    ~=~\delta\G[(p_i\w e^j)\cdot*(p_j\w e^i)\D]&\\
  &=~\G(\delta p_i\w e^j+\delta e^j\w p_i\D)\cdot*\G(p_j\w e^i\D)
    +\G(p_i\w e^j\D)\cdot\delta*\!\G(p_j\w e^i\D)&\\[12pt]
  &e^{klm}\cdot\delta*\!\G(p_j\w e^i\D)
    \stackrel{(\ref{48})}{=}\G(\delta p_j\w e^i
      +\delta e^i\w p_j\D)\cdot*e^{klm}
     -\delta e^{klm}\cdot*\G(p_j\w e^i\D)&
\EEAS
Multiplication by $\F{1}{3!}\GA p_i\w e^j|e_{klm}\DA$ leads to
\BD
  \G(p_i\w e^j\D)\cdot \delta*\!\G(p_j\w e^i\D)=
  \G(\delta p_j\w e^i+\delta e^i\w p_j\D)\cdot*\G(p_i\w e^j\D)
    -\delta e^k\w i_k\!\G(p_i\w e^j\D)\cdot*\G(p_j\w e^i\D)
\ED
and finally one obtains:
\BD
    \delta\G[\GA p_i|*\!e^j\DA\GA p_j|*\!e^i\DA\eta\D]
   =\delta p_i\w 2e^j\GA p_j|*\!e^i\DA
    +\delta e^i\w\G[2p_j\GA p_i|*\!e^j\DA-*e_i\GA p_j|*\!e^k\DA
                       \GA p_k|*\!e^j\DA\D]
\ED
Analogously one finds
\BD
   \delta(p^2\eta)=\delta p_i\w 2pe^i+\delta e^i(2pp_i-p^2*\!e_i)
\ED
and all momentum terms in the equation of motion are proved. The derivation
of the Ricci-scalar term is more difficult:
\BD
  \delta\G(NR_{ij}\w*e^{ij}\D)
    = N\G[\delta\! R_{ij}\w*e^{ij}
      +\delta e^k\w\G(R_{ij}\w*e^{ij}{}_k\D)\D]
\ED
The second term is responsible for the Einstein-force term as shown in
(\ref{9}). It is left to discuss the variation of the curvature-form:
\BEAS
   N\delta\! R_{ij}\w*e^{ij}&=&Nd(\delta\omega_{ij}\w*e^{ij})\\
   &=& d(N\delta\omega_{ij}\w*e^{ij})-dN\w\delta\omega_{ij}\w*e^{ij}\\
   \mbox{\footnotesize modulo exact forms}
      &=&  2\GA \omega_{ij}|e^i\DA  e^j\w *dN
\EEAS
We neglect the exact form which, leads to a surface integral, and substitute
the variation of the connection form by the variation of the triads:
\BEAS
    de^i&=&-\omega^i{}_j\w e^j\\
    \delta de^i&=&-\delta\omega^i{}_j\w e^j-\omega^i{}_j\w\delta e^j\\
    i_i\delta de^i&=&-\GA\delta\omega^i{}_j|e^i\DA e^j-i_i\G(\omega^i{}_j\w
       \delta e^j\D)
\EEAS
Hence we have:
\BEA
\NN   {\TS\F{N}{2}}\delta\!R_{ij}\w*e^{ij}
      &=&\GA\delta\omega_{ij}|e^i\DA e^j\w*dN\\
\NN   &=&-i_i(d\delta e^i+\omega^i{}_j\w \delta e^j)\w*dN\\
\NN   &=&(d\delta e^i+\omega^i{}_j\w\delta e^j)\w*(dN\w e_i)\\
\NN   &=&d(\delta e^i\w*dN\w e_i)+\delta e^i\w D*\!(dN\w e_i)\\
\label{10}
      &=& \delta e^i\w*\G[\nabla\!_idN-e_i\Delta N\D]
\EEA
Of this derivation we keep in mind the last transformation for later use:
\BE
\label{8}
    D*\!(dN\w e_i)=*(\nabla\!_idN-e_i\Delta N)
\EE
Using the equations of motion for each part of the Hamiltonian it is no
difficulty to determine the constraint algebra. All surface integrals are
consequently neglected.
\BE
  \BA{rcl}
     \{H_D(\V N),H_D(\V M)\}&=&H_D([\V N,\V M])\\
     \{H_D(\V N),H_R(a)\}&=&H_R(L_\V Na)\\
     \{H_D(\V N),H_H(N)\}&=&H_H(L_\V NN)\\
     \{H_R(a_1),H_R(a_2)\}&=&-H_R([a_1,a_2])\\
     \{H_R(a),H_H(N)\}&=&0\\
\NN  \{H_H(N),H_H(M)\}&=&H_D(NdM^\sharp-MdN^\sharp)\\
       &&-H_R\G(\GA dN\w dM|e^i{}_j\DA+\GA NdM-MdN|\omega^i{}_j\DA\D)
  \EA
\EE
We calculate only three of these brackets:
\BEAS
    \{H_D(\V N),H_D(\V M)\}&=&\I\G[\F{\delta H_D(\V N)}{\delta e^i}\w
     \F{\delta H_D(\V M)}{\delta p_i} - \F{\delta H_D(\V N)}{\delta p_i}\w
     \F{\delta H_D(\V M)}{\delta e^i}\D]\\
    &=&\I \G[-L_\V Np^i\w L_\V Me_i+L_\V Ne^i\w L_\V Mp_i\D]\\
    &=&\I p^i\w L_{[\V N,\V M]}=H_D([\V N,\V M])
\EEAS
This result reassures that the diffemorphism constraint $H_D$ is the
momentum mapping of the action of the diffeomorphism group.
\BEAS
     \{H_R(a),H_H(N)\}\!&=&\I\Bigg\{a^j{}_ip_j\w N\G(\GA p_k|*e^i\DA e^k
        -{\TS\F{1}{2}}pe^i\D)+a^i{}_j e^j\w\\
     &&\bigg[-N(\GA p_i|*\!e^j\DA p_j-{\TS\F{1}{2}}pp_i)
            +\F{N}{2}(\GA p_l|*\!e^m\DA\GA
            p_m|*\!e^l\DA-{\TS\F{1}{2}}p^2)*\!e_i\\
     &&N*(R_i-{\TS\F{1}{2}}Re_i)+*(\nabla\!_idN-e_i\Delta N
       \bigg]\Bigg\}\\
     &=&\I a^i{}_j\nabla\!_i\nabla^jN\eta=0\qquad,
\EEAS
since $\nabla\!_i\nabla\!_jN$ is symmetric because the connection is
torsionfree.
\BEA
\NN  \{H_H(N),H_H(M)\}&=&\I\bigg[-*\Big(\nabla\!_idN-e_i\Delta N\Big)\w
       M\Big(\GA p_j|*e^i\DA e^j-{\TS\F{1}{2}}p e^i\Big)\\
\NN   &&~~~~+*\!\Big(\nabla\!_i dM-e_i\Delta M\Big)\w
       N\Big(\GA p_j|*e^i\DA e^j-{\TS\F{1}{2}}p e^j\Big)\bigg]\\
\NN   &=&\I \G(N\nabla\!_i\nabla^jM-M\nabla\!_i\nabla^jN\D)p_j\w e^i\\
\NN   &=&\I \G[\nabla\!_i (N\nabla^jM-M\nabla^jN)
          -(\nabla\!_iN\nabla^jM-\nabla\!_iM\nabla^jN)\D]p_j\w e^i\\
\NN   (\ref{5})&=&\I\Bigg\{p_j\w
         L_{(NdM^\sharp-MdN^\sharp)}e^j\\
\NN     &&+\bigg[\omega^j{}_i(NdM^\sharp-MdN^\sharp)+\GA dN\w dM|e^j{}_i\DA
            e^i\w p_j\bigg]\Bigg\}\\
\NN      &=&H_D(NdM^\sharp-MdN^\sharp)\\
\label{11}
        &&-H_R\Big(\GA dN\w dM|e^i{}_j\DA+\GA NdM-MdN|\omega^i{}_j\DA\Big)
\EEA
Here we used the identity
\BE
\label{5}
   DY^i=(\nabla\!_jY^i)e^j=L_\V Ye^i+\omega^i{}_j(\V Y)e^j
\EE
As expected the rotational constraint or more precisely the generated
vectorfields on the infinite dimensional manifold of triads and triad-momenta
form an ideal. So we can go through the canonical analysis without
introducing a fixed coordinate system. One can also define the Poisson
bracket between the 1-form $e^i$ and the 2-form $p_j$:
\BE
   \{e^i(r),p_j(s)\}=\delta^i{}_j\delta(r,s)\qquad r,s\in\S
\EE
Here $\delta$ has to be regarded as a (0,3)-tensorfield on $\S\times\S,~
\delta(r,s)$ is a 1-form with respect to $r$ with values in $\Lambda^2(T_s\S)$
or a 2-form with respect to $s$ with values in $\Lambda^1(T_r\S)$. In a
local chart $(U,\{x^a\}_{a=1,2,3}~\delta(r,s)$ has the following form:
\BE
   \delta(r,s)={\TS\F{1}{2}}\epsilon_{abc}dx^a(s)\w dx^b(s)
      \otimes dx^c(r)\delta(x^1(r)-x^1(s))\delta(x^2(r)-x^2(s))
          \delta(x^3(r)-x^3(s))
\EE
Despite the use of the $\epsilon$-symbol this is a tensor since in coordinate
transformations the product of $\delta$-functions transforms with a
determinant. Finally we give the representation of the usual \ADM momentum
$\pi$, as canonical conjugate to the (0,2)-tensor $q$ a (2,0)-tensorvalued
3-form in terms of the momentum 2-form and the triads
\BE
\label{41}
   \pi = \eta^{k(i}e_i\otimes e_j\otimes p_k\w e^{j)}
       =e_i\otimes e_j p^{(i}\w e^{j)},
\EE
where the symmetrisation is not really necessary, since one can define the
metric momentum only on the constraint surface where $p^i\w e^j$ is
symmetric. This eqution can easily be verified using the well known
relation between $\pi$ and the extrinsic curvature $K=-\omega_{0i}\otimes
e^i$
\BE
   \pi = \Big(K - q \Tr(q^\sharp K)\Big)\otimes \eta,
\EE
and equation (\ref{35}).

\section{The Hilbert-Paltini-action}\RSC
The Hilbert-Palatini-action was suggested by Palatini \cite{Palatini}
mainly in order to avoid second time derivatives of the relevant fields in
the action integral. This can also be achieved by neglecting a surface
integral
{\small
\BE
  {\TS\F{1}{2}}\IM \B R\B \eta\\
     =\IM\bigg[
       \underbrace{-d(\B e^\mu\w*d\B e_\mu)}_{\mbox{\footnotesize boundary
       term}}
       \underbrace{
       -{\TS\F{1}{2}}(d\B e_\mu\w \B e^\nu)\w *(d\B e_\nu\w \B e^\mu)
       +{\TS\F{1}{4}}(d\B e_\mu\w \B e^\mu)\w *(d\B e_\nu\w \B e^\nu)
       }_{\mbox{\footnotesize first order Lagrangian}}\bigg]
\EE}
\hspace{-2mm} but then the action is no longer gauge invariant under
SO(1,3) rotations. Following Palatini's suggestion one regards the action
not only as dependant on the metric, but also on the connection. So the
connection is now no longer torsionfree and metric, since if these
conditions hold, the connection is uniquely determined by the metric.
Unfortunately varying w.r.t. the connection does not yield both necessary
conditions, i.e. if one considers a general linear connection the
Euler-Lagrange equation for the connection only gives a relation between
torsion and non-metricity \cite{schirm}. In the vacuum case this relation
states that the connection is torsionfree, if it is metric, and vice versa.
So one usually assumes that one of the conditions for the Levi-Civit…
connection is satisfied and only varies in the class of either metric or
torsionfree connections. Since in case of a non-metric connection the
compatibility of the gauge group of the tetrades and the connection form is
destroyed -- the connection form for a metric connection is
$so_{\setR}(1,3)$-valued -- we prefer to drop the torsion condition here.
Since our canonical description shall be equivalent to the covariant
formulation, which is of course easier, we first revise the covariant
derivation of the equations. We show that assuming that our connection
satisfies the condition of metricity
\BE
 \B \omega_{\mu\nu}+\B \omega_{\nu\mu}=0
\EE
the Euler-Lagrange equation for the connection form yields in the vacuum
case that our connection is torsionfree.
\BEA
\NN   S(\B e,\B \omega)&=&\F{1}{2}\IM \B R_{\mu\nu}(\B \omega)
             \w*\B e^{\mu\nu}\\
\NN   \delta_\B \omega S&=&\F{1}{2}\IM
\G[d\delta\B \omega_{\mu\nu}\w*\B e^{\mu\nu}
       +\delta\B \omega_{\mu\nu}\w\B \omega^\nu{}_\rho\w*\B e^{\mu\rho}
       +\delta\B \omega_{\mu\nu}\w\B \omega^\mu{}_\rho\w*\B e^{\rho\nu}\D]\\
   &\stackrel{(\ref{26})}{=}&
      \F{1}{2}\IM \delta\B \omega_{\mu\nu}\w \B T^\rho\w*\B
      e^{\mu\nu}{}_\rho
         + d(\delta\B \omega_{\mu\nu}\w*\B e^{\mu\nu})
\EEA
Assuming that the variation vanishs on the boundary  one obtains
\BE
\label{29}
   \F{\delta S}{\delta\B \omega_{\mu\nu}}
    = \B T^\rho\w*\B e^{\mu\nu}{}_\rho=0.
\EE
This turns out to be a zero torsion condition. We show this
abandoning the summation convention for a short while.
\BEAS
  &\DS\sum_{\rho\not=\mu\nu}\B T^\rho\w*\B e^{\mu\nu}{}_\rho=
   \sum_{\rho\not=\mu\nu}*\B e_\rho\GA \B T^\rho|\B e^{\mu\nu}\DA
   +*\B e^\mu\sum_{\rho\not=\mu\nu}\GA \B T^\rho| \B e^\nu{}_\rho\DA
   +*\B e^\nu\sum_{\rho\not=\mu\nu}
   \GA \B T^\rho| \B e_\rho{}^\mu\DA=0&\\[8pt]
  & 1.\qquad\GA \B T^\rho|\B e^{\mu\nu}\DA=0
      \qquad\forall\mu,\nu\not=\rho&\\[4pt]
 &\DS 2.\qquad\sum_{\rho\not=\mu}\GA \B T^\rho|\B e^\nu{}_\rho\DA=0
           \qquad\forall\mu,\nu&
\EEAS
We write $(\GA \B T^\rho|\B e^\nu{}_\rho\DA)_{\rho=1\ld m,\rho\not=\nu}=:\V
x$
and summarize the second equation for fixed $\nu$ for all $\mu$ in the matrix
equation
\BE
  \G(\BA{cccc}
    0&1&\ld&1\\
    \vdots&\ddots&&\vdots\\
    1&\ld&1&0
    \EA\D)\V x=0
\EE
Since this $(m-1)\times(m-1)$-matrix has determinant $(-1)^m(m-2),~\V x=0$
is the only solution if $m \not= 2$ and thus it follows $\B T=0$. Despite
this result we keep in mind that the Palatini-theory is another theory than
Einstein's relativity, since spin fields could now couple to torsion
\cite{Hehl}. As long as there are no experimental data there is only Hehl's
argument, that a theory where only the translatory part of the Poincar'
group couples to geometry seems to be inconsequent. If future experiments
really confirm a spin-torsion-coupling it may well be that one has to modify
the Einstein-Lagrangian, because it does not give rise to derivative terms
for the connection in the field equations. For our canonical description
this means that the there is no time-development of the connection, it is
fully determined by constraints.

The Hamiltonian description we are seeking for should yield the same
result:
The connection to be constructed on the fourdimensional manifold
$\setR\times\S\sim M$ should prove torsionfree. On $\Sigma$ one can only
observe the pullback of the equation $\B T=0$, so it is useful to realise
first the meaning of the pullback of the different components of $\B T$ on
$\S$. We can differ four equations:
\BD
  i_t^*\B T^0=0\qquad i_t^*\B T^i=0\qquad i_t^*i_0\B T^0=0 \qquad
  i_t^*i_0\B T^i=0
\ED
The first equation assures the symmetry of the extrinsic curvature (\ref{6}).
The second implies that the induced connection on $\S$ is torsionfree:
\BE
 0=i_t^*\B T^i
    =i_t^*(d\B e^i+\B \omega^i{}_\mu\w \B e^\mu)=de^i+\omega^i{}_j\w e^j=0
\EE
The third equation yields an important relation between lapse $N$ and
$i_t^*i_0\B \omega^0{}_i$.
\BEAS
    0=i_t^*i_0\B T^0=
     =i_t^*\G[i_0d(\B Nd\B t)+\B e^i  i_0\B \omega^0{}_i\D]
    =-\F{1}{N}dN+e^i i_t^*i_0\B \omega^0{}_i
\EEAS
Defining $b_i:=Ni_t^*\B \omega_{0i}(\B e_0)$ we write this equation as
follows:
\BE
\label{19}
     dN+b_ie^i=0
\EE
The last equation is the well known relation between the extrinsic
curvature, the rotational term and the Lie-derivative along the normal:
\BEA
\NN  &0=i_t^*i_0\B T^i=i_t^*(i_0d\B e^i+\B \omega^i{}_\mu(\B e_0)\B e^\mu
        -\B \omega^i{}_0)
    =i_t^*L_{\B e_0}\B e^i+e^ji_t^*\B \omega^i{}_j(\B e_0)-i_t^*\B
    \omega^i{}_0&\\
     &Ni_t^*L_{\B e_0}\B e^i=-a^i{}_je^j-N\omega_0{}^i&
\EEA
We will now decompose the Lagrange 4-form in another way than in the last
paragraph, since there is a priori no dependance of the connection on the
tetrades.
\BEA
\NN   \hspace{-0.5cm}\B R_{\mu\nu}\w*\B e^{\mu\nu}&=&2\B R_{0i}\w*\B e^{0i}
         +\B R_{ij}\w*\B e^{ij}\\
\NN    &=&2d(\B \omega_{0i}\w*\B e^{0i})+2\B \omega_{0i}\w d\!*\!\B e^{0i}
        +2\B \omega_{0j}\w\B \omega^j{}_i\w*\B e^{0i}\\
\NN      &&+(\,^3\!\B R_{ij}+\B \omega_{i0}\w\B \omega^0{}_j)\w*\B e^{ij}\\
      \mbox{\tiny mod.ex.forms}&\stackrel{(\ref{28})}{=}&
        2\B \omega_{0i}\w d\B e_j\w*\B e^{0ij}+2\B
        \omega_{0j}\w\B \omega^j{}_i\w*\B e^{0i}
         +(\,^3\!\B R_{ij}+\B \omega_{0i}\w\B \omega_{0j})\w*\B e^{ij}~~
\EEA
The pullback $i_t^*$ after insertion of the time vector field $\P/\P\B t$
will be simple, if we first eliminate $\B e^0$ from the last factor
$\B e^{\ld}$. Then the dualised form $*\B e^{\ld}$ will contain a factor
$\B e^0=\B Nd\B t$ in any case and only if the time vector field is
inserted there the term will not vanish after the pull back to $\S$.
Technically spoken $i_t^*i_{\P/\P\B t}$ is performed then by changing
$\stackrel{\SSS 4}{*}$ to $\stackrel{\SSS 3}{*}$ and multiplying with the
lapse function.
\BEA
\NN  \B{\C L}&=&\F{1}{2}\G[2\B \omega_{0i}\w d\B e_j\w i^0\!*\!\B e^{ij}
       -2\B \omega_{0j}\w\B \omega^j{}_i\w i^0\!*\!\B e^i
       +(\,^3\!\B R_{ij}+\B \omega_{0i}\w\B \omega_{0j})\w*\B e^{ij}\D]\\
\NN  &=&-\B \omega_{0i}(\B e_0)d\B e_j\w*\B e^{ij}+\B \omega_{0i}\w
           i_0d\B e_j\w*\B e^{ij}
       -\B \omega_{0i}(\B e_0)\B \omega^i{}_j\w*\B e^j
       -\B \omega^i{}_j(\B e_0)\B \omega_{0i}\w*\B e^j\\
     &&+{\TS\F{1}{2}}(\,^3\!\B R_{ij}+\B \omega_{0i}\w\B \omega_{0j})
     \w*\B e^{ij}
\EEA
Using the definitions $b_i:=Ni_t^*\B \omega_{0i}(\B e_0)$ and
$a^i{}_j:=Ni_t^*\B \omega^i{}_j(\B e_0)$
we obtain:
\BEAS
  \C L&=&i_t^*i_{\P/\P\B t}\B{\C L}\\
     &=&\G[{\TS\F{N}{2}}(R_{ij}+ \omega_{0i}\w\omega_{0j})
          -b_ide_j+\omega_{0i}\w Ni_t^*i_0d\B e_j\D]
             \w\stackrel{3}{*}\!e^{ij}
          -\G(b_i\omega^i{}_j-a^i{}_j\omega_{0i}\D)\w\stackrel{3}{*}\!e^{j}\\
    &=&\G[{\TS\F{N}{2}}(R_{ij}+\omega_{0i}\w\omega_{0j})-
       b_i(de_j+\omega_{jk}\w e^k)+a^k{}_je_k\w\omega_{0i}
         +\omega_{0i}\w Ni_t^*i_0d\B e_j\D]\w*e^{ij}
\EEAS
In order to introduce $\dot{e}^i$ we again have to use the space-time
picture:
\BE
    \dot{e}^i=i_t^*L_{\P/\P\B t}\B e^i=L_\V Ne^i+Ni_t^*L_{\B e_0}\B e^i
       =L_\V Ne^i+Ni_t^*i_0d\B e^i
\EE
Then we obtain the following Lagrangian:
\BEA
\NN  &&L(e^i,\dot{e}^i,\omega_{0i},\omega_{ij},N,\V
N,a^i{}_j,b_i)\\
\NN     &=&\I \bigg\{\dot{e}^i\w\omega_{0j}\w*e_i{}^j
        -\Big[-{\TS\F{N}{2}}(R_{ij}+\omega_{0i}\w\omega_{0j})\\
       &&~~~~+b_i(de_j+\omega_{jk}\w e^k)+a^k{}_ie_k\w\omega_{0j}
                  +L_\V Ne_i\w\omega_{0j}\Big]\w*e^{ij}\bigg\}
\EEA
We notice that the time derivatives of the connection form
components $\omega_{ij},~\omega_{0i},~b_i$ and $a^i{}_j$ do
not appear  in the Lagrangian like those of lapse and shift.
This is not surprising, since in the covariant equations
$\F{\delta S}{\delta\omega_{\mu\nu}}=0$ did not contain an
exterior derivative of the connection form. We can reduce
the number of constraints that we get, if we identify at
this point
\BE
   \F{\delta L}{\delta\dot{e}^i}=p_{e^i}=\omega_{0j}\w*e_i{}^j\qquad.
\EE
Formally spoken this identification solves the pair of second class
constraints
\BE
    p_{e^i}-\omega_{0j}\w*e_i{}^j\approx 0
            \qquad p_{\omega_{0i}}\approx 0\qquad.
\EE
The spatial part of the connection form $\B \omega_{0i}$ is then already
eliminated as a redundant degree of freedom. Before this identification one
would obtain the equation $L_{Nn}e^i+a^i{}_j-N\omega^i{}_0=0$ as
nondynamical Euler-Lagrange-equation for $\omega_{0i}$. After the
identification we obtain this equation in the Hamiltonian picture as
equation of motion for $\dot{e}^i$. The Hamiltonian is now easily obtained
($p_{e^i}\equiv p_i$):
\BEA
\NN   &&H(e^i,p_i,\omega_{ij},N,\V N,a^i{}_j,b_i)=\\
\NN     &=&\I p_i\w L_\V Ne^i-a^i{}_je^j\w p_i
        +b_i(de_j+\omega_{jk}\w e^k)\w*e^{ij}\\
   &&~~~ +{\TS\F{N}{2}}\G(\GA p_i|*\!e^j\DA
              \GA p_j|*\!e^i\DA -{\TS\F{1}{2}}p^2-R\D)\eta
\EEA
Compared to the \ADM Hamiltonian there is one additional term. Since
$de^j+\omega^j{}_k\w e^k$ is the torsion of the connection we have an explicit
torsion potential. Apart from the two constraints which were already
discussed and solved the Lagrangian gives rise to five other primary
constraints:
\BE
     p_N\approx 0\qquad p_\V N\approx 0\qquad p_{a_{ij}}\approx 0\qquad
     p_{b_i}\approx 0\qquad p_{\omega_{ij}}\approx 0
\EE
These constraints give rise to the following secondary constraints:
\BE
 \BA{rcccll}
   C_H&=&\{H,p_N\}&=&{\TS\F{1}{2}}\G(\GA p_i|*\!e^j\DA\GA p_j|*\!e^i\DA
      -{\TS\F{1}{2}}p^2-R\D)\eta&\mbox{Hamiltonian}\\
   {C_D}_i&=&\{H,p_{N^i}\}&=&-dp_i +p_j\w
   i_ide^j&\mbox{diffeomorphism}\\
\hspace{-5mm}{C_R}^{ij}&=&\{H,p_{a_{ij}}\}
         &=&-{\TS\F{1}{2}}(p^i\w e^j-p^j\w e^i)&\mbox{rotational}\\
   {C_T}^i&=&\{H,p_{b^i}\}&=&(de^j+\omega^j{}_k\w e^k)\w*e^i{}_j
        =T^j\w*e^i{}_j&\mbox{torsion}\\
   {C_C}^{ij}&=&\{H,p_{\omega_{ij}}\}&=&-(dN+b_k e^k)\w*e^{ij}
         -NT^k\w*e^{ij}{}_k&\mbox{connection constraint}
  \EA
\EE
The Hamilton function is again a sum of integrated secondary constraints.
But only four of the five secondary constraints appear in the Hamiltonian.
\BD
 \BA{cccccccl}
  &\multicolumn{7}{l}{H(e^i,p_i,N,\V N,\omega_{ij},a_{ij},b_i)}\\[0.1cm]
  =& \I N^i{C_D}_i(e,p)&+&\I a_{ij}{C_R}^{ij}(e,p)&
 +&\I C_H(e,p,\omega)&+&\I b_i{C_T}^i(e,\omega)\\[0.1cm]
 =&H_D(\vec{N};e,p)&+&H_R(B;e,p)&+&H_H(N;e,p,\omega)&+&H_T(A_i;e,\omega)
 \EA
\ED
We observe that the rotational, torsion and connection constraint imply the
remaining equations which assure that the connection to be constructed on
$\setR\times\S$ is torsionfree: Because of the rotational constraint the
extrinsic curvature is symmetric. Note that this is a secondary
constraint here whereas it is a primary constraint in the \ADM
description.
The other equations are found in the following way:
\BEA
\NN  &0\approx {C_C}^{ij}\w e_j= 2(dN+b_k\w e^k)\w*e^i+NT^j\w*e^i{}_j
       \approx 2(dN+b_ke^k)\w*e^i&\\[16pt]
     &\Lra\qquad dN+b_ke^k\approx 0&
\EEA
because of the torsion constraint and thus both terms in this sum for $C_C$
vanish separately, hence
\BE
    T^k\w*e^{ij}{}_k\approx 0
\EE
and this equation implies $T^k\approx 0$ as shown above (\ref{29}). Note
that these arguments are independent of the space dimension. In order to
simplify the constraint analysis we substitute  torsion and connection
constraint by the following equivalents:
\BEA
    C_1^i&:=&de^i+\omega^i{}_j\w e^j\approx 0\\
    C_2  &:=&dN+b_ide^i\approx 0
\EEA
We do not expect any tertiary constraints since there is no equivalent on
the covariant level, but it is not obvious that there exists an extension
of the Hamiltonian by a sum of integrated primary constraints which does
conserve all constraints. To check the absence of tertiary constraint it is
necessary to find integration functions $K_1,K_2,K_3,K_4,\kappa_5$ such
that
\BE
  \{H+\I K_1p_N+\I {K_2}^ip_{N^i}+\I {K_3}_{ij}p_{a_{ij}}+\I {K_4}_ip_{b_i}
     +\I {\kappa_5}_{ij}\w p_{\omega_{ij}},C_{H/D/R/1/2}\}\approx 0
\EE
We will first analyze the constraints because then it will become obvious
how to choose these integration variables. For definition and a survey we
summarize once again all constraints we have found. One should note the
difference of test integration forms and canonical variables:

{\footnotesize
\BD
\renewcommand{\arraystretch}{1.6}
 \arraycolsep0.9mm\BA{c|rcl|rcl}
  &\multicolumn{3}{c|}{\mbox{primary}}&\multicolumn{3}{c}{\mbox{secondary}}\\
  \hline
    & H_3(\rho_{ij};p_{\omega_{ij}})
     &:=&{\TS\F{1}{2}}\I\rho_{ij}\w p_{\omega_{ij}}&
       H_1(\alpha_i;e^i,\omega_{ij})
     &:=&\I\alpha_i\w(de^i+\omega^i{}_j\w e^j)\\
\mbox{second}&\rho_{ij}=-\rho_{ji}&\in&\Omega^1(\S)
       &\alpha_i&\in&\Omega^1(\S)\\
\mbox{class}&H_4(R_i;p_{b_i})&:=&\I R_ip_{b_i}&
       H_2(\beta;e^i,N,b_i)&:=&\I\beta\w(dN+b_ie^i)\\
     &  R_i&\in&\C C^\infty(\S)&\beta&\in&\Omega^2(\S)\\
  \hline
\mbox{first}&H_N(U;p_N)&:=&\I Up_N&
      H_H(X;e^i,p_i,\omega_{ij})&=&\I{\TS\F{X}{2}}
         [\GA p_i|*\!e^j\DA\GA p_j|*\!e^i\DA-{\TS\F{1}{2}}p^2-R]\eta\\
\mbox{class}&H_\V N(V^i;p_{\V N})&:=&\I V^ip_{N^i}&
        H_D(\V Y;e^i,p_i)&=&\I p_i\w L_\V Ne^i\\
        &H_a(W_{ij};p_{a_{ij}})&:=&\I W_{ij}p_{a_{ij}}&
        H_R(Z_{ij};e^i,p_i)&:=&-\I Z_{ij}p^i\w e^j\qquad Z_{ij}=-Z_{ji}
  \EA
\ED}
\hspace{-1.9mm}{\bf Theorem:}\\
The constraints $H_1,\ld,H_4$ form second class pairs. The other constraints
$H_N,~H_\V N,\linebreak H_a,H_H,~H_D,~H_R$ can be substituted by
equivalent  constraints $H_\C N,~H_\V{\C N},~H_\C A,~H_\C H,\linebreak
H_\C D,H_\C R$ which are first class. The Poisson bracket relations of
those substituted constraints correspond to the relations in the \ADM
formulation and since they are first class, they also equal the Dirac
brackets of the original constraints which one could have calculated
directly as well. All surface terms are consequently neglected.

\pr\\
The constraints $H_1,\ld,H_4$ are obviously second class, one obtains for
$(\{H_i,H_j\})_{ij=1\ld 4}$ a skewsymmetric matrix of functionals:
{\small
\BE
(\{H_i,H_j\})_{ij=1\ld 4}=\!\!\G({\arraycolsep-.3mm
    \BA{cccc}
      0&0&-\I\rho_{ij}\w\alpha^i\w e^j&0\\
      0&0&0&\I R_i\beta\w e^i\\
      \I\rho_{ij}\w\alpha^i\w e^j&0&0&0\\
      0&-\I R_i\beta\w e^i&0&0
    \EA}\,\D)=:(S_{ij})_{ij=1\ld 4}
\EE}
\hspace{-2mm} This matrix is invertible and thus the matrix of Poisson
brackets of all constraints has at least rank 4 and because the constraints
$H_1,\ld,H_4$ yield this simple symplectic structure we regard these
constraints as the fundamental second class pairs. We know that on the
surface described by these second class constraints the induced connection
is torsionfree, but the extrinsic curvature is not necessarily symmetric.
We suppose that all other constraints are first class, since there should
not be less first class constraints than in the \ADM  case. Starting with
the primary constraints we can set
\BE
   H_\V{\C N}:=H_\V N\qquad\qquad H_\C A:=H_a
\EE
since these constraints are first class. $H_N$ does not commute with $H_2$
because of the lapse dependance. Thus we substitute $H_N$ in the following
way:
\BEA
\NN  H_\C N(U;p_N,p_{b_i})&:=&H_N(U;e^i,p_i)
        -H_4(\GA dU|e_i\DA;p_{b_i})\\
        &=&\I \Big(Up_N-\GA dU|e_i\DA p_{b_i}\Big)
\EEA
One can easily check that this constraint commutes also with $H_2$. Regarding
the secondary constraints we notice that none of them commutes with $H_1$
and $H_2$ because of their momentum dependance. But the Hamiltonian
constraint $H_H$ does not even commute with the primary constraint $H_3$.
And indeed calculating for example the Poisson bracket
\BE
    \{H_H(N),H_H(M)\}=0
\EE
we obtain a result which differs from the \ADM  result, because here the
Ricci-scalar-term remains underived since the connection form $\omega$ does
not depend on the triads $e$ before restriction to the constraint surface
and thus $\F{\delta H(N)}{\delta e^i}$ and $\F{\delta H(N)}{\delta p_i}$
are both linear in $N$. In order to make the Hamiltonian constraint $H_H$
commute with $H_3$ we have to add a term $H_1$ with a certain argument
which we now determine:
\BEAS
   &\T H_H(X):=H_H(X)-H_1(\alpha_i(X))&\\[4pt]
   &\{\T H_H(X),H_3(\rho_{ij})\}\simeq\!\footnotemark~0&
\EEAS
\footnotetext{$\simeq$ denotes: after restriction to the second class
surface}
\vspace{-20pt}
\BEAS
  \{H_H(X),H_3(\rho_{ij})\}&=&\I\G[-d(X*\!e^{ij})-X\omega^j{}_k\w*e^{ik}
    -X\omega^i{}_k\w*e^{kj}\D]\w{\TS\F{1}{2}}\rho_{ij}\\
  &\stackrel{(\ref{26})}{\simeq}&-{\TS\F{1}{2}}\I\rho_{ij}\w dX\w*e^{ij}\\
  \{H_1(\alpha_i),H_3(\rho_{ij})\}&=&-\I\rho_{ij}\w\alpha^i\w e^j
\EEAS
Thus we obtain $\alpha^i\w e^j-\alpha^j\w e^i=+dX\w*e^{ij}$ or equivalently:
\BD
    \alpha^i\w*e_{ki}=+dX\w e_k
\ED
In the same way as we transformed $p_i$ to $\omega_{0i}$ in (\ref{7})
we obtain
\BD
   \alpha^i=-\GA dX\w e_k|*\!e^i\DA e^k
      +{\TS\F{1}{2}}\GA dX \w e_j|*\!e^j\DA e^i=*(dX\w e^i)
\ED
and our corrected Hamiltonian constraint $\T H$ reads:
\BE
  \T H(X):=H_H(X)-H_1(*(dX\w e_i))\footnotemark
\EE
\footnotetext
{The correction term seems to be a pure surface integral
\BEAS
   H_1(*(dX\w e_i))&=&\TS\I *(dX\w e_i)\w (de^i+\omega^i{}_j\w e^j)\\
   &=&\TS\I D*(dX\w e_i)\w e^i-\I d\G[*(dX\w e_i)\w e^i\D]\\
   &\stackrel{(\ref{8})}{=}&
     \TS\I *(\nabla\!_idX-e_i\Delta X)\w e^i-\int_{\P\S}\ld\\
   &=&-2\TS\I\nabla\!_i\nabla^iX\eta,
\EEAS
 but Gauss's law does not hold for connections with torsion:
\BD
   \nabla\!_i\nabla^iX\eta\not=L_{(e_i\cdot\nabla^iX)}\eta
           =d(\nabla\!_iX\cdot\w*e^i)
\ED
}
$\T H,~H_D$ and $H_R$ do now commute with all primary constraints. Forming
the Poisson brackets among themselves we are already back to the \ADM
algebra if one restricts to the second class constraint surface after
calculation of the brackets. We will show this for the two most difficult
examples:
{\arraycolsep 1.4mm
\BEA
\NN &&\{\T H_H(X),\T H_H(Y)\}\\
\NN    &=&-\{H_1(*(dX\w e^i)),H_H(Y)\}
                          -\{H_H(X),H_1(*(dY\w e_i))\}\\
\NN &=&-\I\frac{\delta}{\delta e^i}\G[*(dX\w e_k)\w
       (de^k+\omega^k{}_j\w e^j)\D]\w
       Y\G(\GA p_j|*\!e^i\DA e^j-{\TS\frac{1}{2}}pe^i\D)
       +(X\longleftrightarrow Y)\\
\NN &=&-\I \G[d*(dX\w e_i)-\omega^k{}_i\w*(dX\w e_k)\D]\w
       Y\G(\GA p_j|*\!e^i\DA e^j-{\TS\frac{1}{2}}pe^i\D)
        +(X\longleftrightarrow Y)\\
\NN &&-\I \G[\F{\delta}{\delta e^i}*(dX\w e_k)\D]
       \G(de^k+\omega^k{}_j\w e^j\D)\w
       Y\G(\GA p_j|*\!e^i\DA e^j-{\TS\frac{1}{2}}pe^i\D)
        +(X\longleftrightarrow Y)\\
\NN &\simeq&\I -*\G(\nabla\!_idX-e_i\Delta X\D)\w
              Y\G(\GA p_j|*\!e^i\DA e^j-{\TS\frac{1}{2}}p e^i\D)
        +(X\longleftrightarrow Y)\\
  &\stackrel{\mbox{\tiny (\ref{11})}}{=}&H_D(XdY^\sharp-YdX^\sharp)
   -H_R\Big(\GA dX\w dY|e^i{}_j\DA
   +\GA XdY-YdX| \omega^i{}_j\Big)
\EEA}
In order to calculate the bracket $\{H_D(\V Y),\T H_H(X)\}$ we split $H_H(X)$
in the parts
\BEA
   {H_H}_{\SSS kin}(X;e,p)&:=&
      {\TS\F{1}{2}}\I X\G(\GA p_i|*\!e^j\DA
      \GA p_j|*\!e^i\DA-{\TS\F{1}{2}}p^2\D)\\
   {H_H}_{\SSS pot}(X;e,p,\omega)&:=&-{\TS\F{1}{2}}\I XR_{ij}(\omega)\w*e^{ij}
\EEA
and obtain:
\BEAS
  \{H_D(\V Y),{H_H}_{\SSS kin}(X)\}&=&{H_H}_{\SSS kin}(L_\V YX)\\
  \{H_D(\V Y),{H_H}_{\SSS pot}(X)\}&=&
    {\TS\F{1}{2}}\I L_\V Ye^i\w XR_{jk}(\omega)*\!e^{jk}{}_i\\
    &\simeq&{\TS\F{1}{2}}\I L_\V Y*\!e^{jk}\w XR_{jk}^{LC}(e)\\
    &=&{\TS\F{1}{2}}\I \bigg[L_\V Y(X*\!e^{jk}\w R_{jk})\\
    &&~~~~-(L_\V YX)\big(R_{jk}\w*e^{jk}\big)
       -X*e^{jk}\w L_\V YR_{jk}(e)\bigg]\\
    \mbox{\tiny mod.bound.terms.}&\stackrel{(\ref{10})}{=}&
   {H_H}_{\SSS pot}(L_\V YX)-\I L_\V Ye^j\w *(\nabla\!_jdX-e_j\Delta X)\\
    &\simeq&{H_H}_{\SSS pot}(L_\V YX)+\{H_D(\V Y),H_1(*(dX\w e^i))\}
\EEAS
Thus we have
\BE
  \{H_D(\V Y),\T H(X)\}\simeq H_H(L_\V YX)\simeq \T H_H(L_\V YX)
\EE
since the correction term vanishes on the second class constraint surface.

In the last mainly technical step we must change $\T H_H,~H_D$ and $H_R$ in
such a way that they also commute with $H_1$ and $H_2$, the secondary second
class constraints. It is obvious that one has to add $H_3$- and $H_4$-terms
with certain integration functions. We just show this for the case of the
rotational constraint:
\BEAS
  \{H_R(Z_{ij}),H_1(\alpha^i)\}&=&\I Z_{ij}e^j\w D\alpha^i
    \stackrel{m.e.f.}{\simeq}\I DZ_{ij}\w \alpha^i\w e^j\\
  \{H_R(Z_{ij}),H_2(\beta)\}&=&\I Z_{ij}b^i\beta\w e^j
\EEAS
Thus we substitute $H_R$ in the following way:
\BE
  H_\C R(Z_{ij}):=H_R(Z_{ij})
         +K_R(Z_{ij};\omega_{ij},b_i,p_{\omega_{ij}},p_{b_i})
    :=H_R(Z_{ij})-H_3(DZ_{ij})-H_4(Z_{ij}b^j)
\EE
In the same way one can find correction terms for the diffeomorphism and the
Hamiltonian constraint $H_D$ and $\T H_H$:
\BEA
\NN  &H_\C D(\V Y):=H_D(\V Y)+K_D(\V Y)\qquad\qquad
     K_D(\V Y)=K_D(\V Y;e,\omega,b,N,p_\omega,p_b)&\\[2pt]
     &K_D(\V Y):=H_3(L_\V Y\omega_{ij})+H_4(b_j\GA L_\V Ye^j|e_i\DA)&\\[4pt]
\NN    &H_\C H(X):=\T H(X)+K_H(X)\qquad\qquad
     K_H(X)=K_H(X;e,p,\omega,b,N,p_\omega,p_b)&\\[2pt]
     &K_H(X):=-H_3\G({\TS\F{1}{2}}[i_i\beta_j-i_j\beta_i+e^ki_{ij}\beta_k]\D)
           -H_4\G(b_i(X\GA p_j|*\! e^i\DA-{\TS\F{1}{2}}p\eta^i{}_j)\D)&\\
\NN        &\beta_i:=D\G[X\GA p_k|*\!e_i\DA e^k-{\TS\F{1}{2}}pe_i\D]&
\EEA
The Hamiltonian constraint is quadratic in the momentum $p$, and thus the
correction term depends on $p$. One should convince oneself that despite of
the dependance of the integration functions on the canonical variables the
Poisson brackets restricted to the second class constraint surface are
unchanged. Using Leibniz's rule for deriving the correction terms the
derivation of the integration function yields in any case the constraint
functional which vanishs after restriction to the constraint surface.

We will show now at an example that having proved that there are
exactly two pairs of second class constraints one could also have calculated
the Dirac brackets of the original constraints in order to find the Poisson
bracket relations of the corrected first class constraints:
\BEAS
 \{H_H(X),H_H(Y)\}_D&=&\{H_H(X),H_H(Y)\}
       -\{H_H(X),H_i(\cdot)\}S^{ij}(\cdot,\cdot)\{H_j(\cdot),H_H(Y)\}\\
    &=&-\{H_H(X),H_1(\alpha)\}S^{13}(\alpha,\rho)\{H_3(\rho),H_H(Y)\}\\
    &&-\{H_H(X),H_3(\rho')\}S^{31}(\alpha',\rho')\{H_1(\alpha'),H_H(Y)\}
\EEAS
where $S^{ij}$ is the inverse of the matrix of second class constraints. This
expression should be independent of the chosen integration forms $\alpha$
and $\rho$, and so we choose \linebreak $\alpha$ depending on $Y$ resp.
$\alpha'$ depending on $X$, such that the product $S^{13}(\alpha,\rho)\cdot
\linebreak[4]\{H_3(\rho),H_H(Y)\}$ and $\{H_H(X),H_3(\rho')\}
S^{31}(\alpha',\rho')$ is 1.The calculation will be equivalent to the
determination of the correction term for $H_H$ to $\T H_H$:
\BEAS
  S^{13}(\alpha,\rho)&=&\G(\I \rho_{ij}\w\alpha^i\w e^j\D)^{-1}\\
  \{H_3(\rho),H_H(Y)\}&=&{\TS\F{1}{2}}\I\rho_{ij}\w dY\w*e^{ij}
     =\I \GA\rho_{ij}|e^i\DA\GA dY|e^j\DA\eta\\
     &=&\I\rho_{ij}\w*(dY\w e^i)\w e^j
\EEAS
So one has to choose $\alpha^i=*(dY\w e^i)$ and $\alpha'^i=*(dX\w e^i)$
and the calculation proceeds as shown in (\ref{11}):
\BEAS
\NN  \{H_H(X),H_H(Y)\}_D&\!\!=&\!\!-\{H_H(X),H_1\G(*(dY\w e^i)\D)\}+
      \{H_H(Y),H_1\G(*(dX\w e^i)\D)\}\\
     &\!\!\simeq& \!\!H_D(XdY^\sharp\!-\!YdX^\sharp)-H_R\Big(\!\GA dX\!\w\!
     dY|e^i{}_j\DA\!+\GA XdY\!-\!YdX|\omega^i{}_j\DA\! \Big)
\EEAS
One could as well have chosen $\rho$ depending on $X$ and $\rho'$ depending
on $Y$, such that $\{H_H(X),H_1(\alpha)\}S^{13}(\alpha,\rho)$ and
$S^{31}(\alpha',\rho')\{H_H(\alpha'),H(Y)\}$ is 1. This calculation is
equivalent to the determination of the correction term for $\T H_H$ to
$H_\C H$.

Finally we have to show that there is an extension of our Hamiltonian by
primary constraints which conserves all constraints. Since our Hamiltonian is
a sum of constraints:
\BEAS
   H(e,p,N,\V N,\omega,a,b)&=&H_H(N)+H_1(*(b_je^{ji}))+H_D(\V N)+H_R(a)\\
   &\simeq&\T H_H(N)+H_D(\V N)+H_R(a)
\EEAS
and we know how to correct these constraints in order to get first class
constraints we suppose that
\BEA
\NN  \C H&:=&\underbrace{H(e,p,N,\V N,\omega,a,b)+K_H(N)+K_D(\V N)+K_R(a)}_
        {\mbox{\footnotesize special first class Hamiltonian}}\\
   &&+\underbrace{H_\C N(U)+H_{\V{\C N}}(\V V)+H_\C A(W)}_
         {\mbox{\footnotesize arbitrary primary first class term}}
\EEA
is a first class Hamiltonian. This is actually the most general first class
Hamiltonian for our infinite-dimensional problem  (\cite{Dirac}(1.32)).
To prove that the first part is first class, we rewrite it in the
following way:
\BEA
\NN    \C H&=&H(e,p,N,\V N,\omega,a,b)+K_H(N)+K_D(\V N)+K_R(a)\\
        &=&H_H(N)+H_1(*(b_je^{ji}))+K_H(N)+H_\C D(\V N)+H_\C R(a)
\EEA
Forming Poisson brackets with primary constraints yield secondary
constraints, forming Poisson brackets with secondary constraints yield
secondary constraints, since one obtains the same result as if one had taken
the bracket with
\BD
  H_\C H(N)+H_\C D(\V N)+H_\C R(a)
    =H_H(N)-H_1(*(dN\w e^i))+K_H(N)+H_\C D(\V N)+H_\C R(a)
\ED
apart from a term $H_1(\cdot)$ which is a result of the different
derivations of the integrand functions of the $H_1$ term and which vanishes
on the surface described by the second class constraints.

Thus we have shown that three of the four different equations which represent
in 3+1 dimensions that the connection on the fourdimensional manifold is
torsionfree are realised by second class constraints, whereas the fourth is
obtained by a first class constraint. If one restricts to the second class
constraint surface from the beginning the only difference to the \ADM
formulation is that there is one more secondary constraint and that the part
of the phase space where the rotational constraint does not equal zero can be
interpreted as a region, where the zero component of the torsion does not
vanish. As a by-product we have seen how Dirac's framework for second class
constraints works for a field theory when the canonical variables are forms
rather than functions.

\section{The Ashtekar formulation} \RSC
It is well known the main idea of the Ashtekar formulation is the
restriction to four dimensions and the use of a complex selfdual
connection \cite{Ashb}. We will show that the effect is a canonical
transformation in the complexified phase space. In order to obtain a
simplification it is necessary to perform a second transformation which is
normally done by the use of densities. We are free to decide if we want to
regard the selfdual connection as an independent variable or if we prefer to
work with the unique selfdual connection given by the selfdual projection of
the Levi-Civit…-connection. Since we know that the Palatini theory reduces
to the \ADM  theory in the vacuum case -- otherwise spin could couple with
torsion \cite{Hehl} -- we attempt a formulation where the four-dimensional
connection is the complex selfdual projection of the Levi-Civit…-connection.
It is no contradiction that the connection form of the spatial restriction
of the connection turns out to be one of the canonical variables, as in the
\ADM  theory it is no contradiction that the momentum consists mainly of one
component $\omega_{0i}$ of the connection. Before describing the canonical
formulation we shortly summarize some selfdual notation. Our first step is
the transition to the complexified tangential bundle over the real
space-time manifold $M$. We consider the complexified frame bundle
$L_C(M)$ which carries a certain real structure, because we still regard
the manifold as real. We will regard the complexified bundle of tetrades
where the gauge group is now $SO_C(1,3)$ as a subbundle of the tangential
bundle. As usual there is a unique metric torsionfree connection whose
components
\BE
 \B \omega_{\alpha\, \beta\gamma}={\TS\F{1}{2}}\G[\B g_{\alpha\beta,\gamma}
       -\B g_{\beta\gamma,\alpha}+\B g_{\gamma\alpha,\beta}
       +\B C_{\alpha\,\beta\gamma}-\B C_{\beta\,\gamma\alpha}
       +\B C_{\gamma\,\alpha\beta}\D]
\EE
are real only if they belong to a real basis, for example a coordinate
basis.

In the Lie-algebra $so_{\setR}(1,3)$ one can define the dualization by
\BE
  A_{\rho\sigma}:={\TS\F{1}{2}}\epsilon_{\rho\sigma}{}^{\mu\nu}A_{\mu\nu}
\EE
Because of the signature of the metric one finds
\BE
  \widetilde{\T A}=-A
\EE
so the eigenvalues of the dualisation operator are $\pm i$ and hence a
decomposition in eigenvectors is only possible in the complexified algebra
$so_C(1,3)$. One calls elements of $so_C(1,3)$ with the property
\vspace{-12pt}
\BEAS
   &\T A=iA&\qquad\qquad\mbox{selfdual}\\
   \mbox{and}\qquad\qquad&\T
   A=-iA&\qquad\qquad\mbox{antiselfdual.}
\EEAS
The subspaces
\vspace{-24pt}
\BEAS
  {}\qquad so_C^+(1,3)&:=&\{A\in so_C(1,3)|\T A=iA\}\qquad\qquad\mbox{and}\\
  so_C^-(1,3)&:=&\{A\in so_C(1,3)|\T A=-iA\}
\EEAS
form ideals and it holds
\BD
    so_C(1,3)=so_C^+(1,3)\oplus so_C^-(1,3)\qquad.
\ED
For a proof one shows
\BE
\label{12}
   \widetilde{[A,B]}=[A,\T B]=[\T A,B]
\EE
One can introduce the projectors to the selfdual and antiselfdual parts
\BE
   P^\pm A:={\TS\F{1}{2}}(A\mp i\T A) \qquad\qquad P^\pm A=:A^\pm
\EE
and with help of (\ref{12}) one can directly show
\BE
   P^\pm[A,B]=[P^\pm A,B]=[A,P^\pm B],
\EE
which proves that $so_C^\pm(1,3)$ form ideals of $so_C(1,3)$.
The connection form with respect to a tetrad base is $so_C(1,3)$-valued, and
the associated curvature form is a $so_C(1,3)$-valued  2-form. Since
\BE
   \B R^++\B R^-=\B R= d\B \omega+\B \omega\w\B \omega=
     \underbrace{d\B \omega^++\B \omega^+\w\B \omega^+}_{\mbox{\footnotesize
     selfdual}}
    +\underbrace{d\B \omega^-+\B \omega^-\w\B \omega^-}_{\mbox{\footnotesize
      anti-selfdual}}+
   \underbrace{[\B \omega^+\stackrel{\w}{,}\B
   \omega^-]}_{\mbox{\footnotesize
   0}}
\EE
the (anti-) selfdual part of the curvature is the curvature to the (anti-)
selfdual part of the connection form. We now return to the action principle.
Using the relation
\BE
   \B R_{\mu\nu}\w \B e^\nu=0
\EE
we can consider the selfdual Einstein-Hilbert action instead of the
usual one.
\BE
  S=\I \B R^{\SSS +}_{\mu\nu}\w *\B e^{\mu\nu}
   =\F{1}{2}\I (\B R_{\mu\nu}\w*\B e^{\mu\nu}-i\B R_{\rho\sigma}\w
   \B e^{\rho\sigma})=\F{1}{2}\I \B R_{\mu\nu}\w*\B e^{\mu\nu}
\EE
Varying with respect to the tetrades yields (the complexified version of)
Einstein's equations. In order to obtain the Hamiltonian formulation we
decompose the Lagrange form ($\epsilon_{0ijk}\equiv\epsilon_{ijk}$):
\vspace{-24pt}
\BEA
\NN  \B{\C L}&=&{\B R^{\SSS +}}_{\mu\nu}\w*\B e^{\mu\nu}\\
\NN   &=&2\B R^{\SSS +}_{0i}\w*\B e^{0i}+\B R^{\SSS +}_{ij}\w*\B e^{ij}\\
\NN   &=&-i\epsilon_{ijk}\B R^{{\SSS +}\,jk}\w*\B e^{0i}
         +\B R^{\SSS +}_{ij}\w*\B e^{ij}\\
\NN   &=&+i\epsilon_{ijk}i_0\B R^{{\SSS +}\,jk}\w*\B e^i
     +\B R^{\SSS +}_{ij}\w*\B e^{ij}\\
     &=&d\B t\w[\epsilon_{ijk}\B Ni_0\B R^{{\SSS +}\,jk}
     \w\stackrel{3}{*}\!\B e^i
       +\B N\B R^{\SSS +}_{ij}\w\stackrel{3}{*}\!\B e^{ij}]
\EEA
Now we use the following equation
\BEA
\NN   \B Ni_0\B R^{{\SSS +}\,jk}&=&\B Ni_0\G(d\omegap^{jk}
        +\omegap^j{}_\mu\w\omegap^{\mu k}\D)
        =i_{\B N\B e_0}d\omegap^{jk}+di_{\B N\B e_0}\omegap^{jk}
          -\,^\omegap\!\!D(i_{\B N\B e_0}\omegap^{jk})\\
        &=&L_{\B N \B e_0}d\omegap^{jk}
        -\,^\omegap\!\!Di_{\B N \B e_0}
        \omegap^{jk},
\EEA
where $L$ denotes the Lie-derivative and obtain
\BE
      \B{\C L}=d\B t\w\G[iL_{\P/\P\B t}\omegap_{jk}\w \B e^{jk}
        -iL_{\B{\V N}}\omegap_{jk}\w
       \B e^{jk}-i\,^\omegap\!\!Di_{\B N \B e_0}\omegap_{jk}
       \w \B e^{jk}+N\B R^{\SSS +}_{ij}\w\stackrel{3}{*}\! \B e^{ij}\D].
\EE
Since
\vspace{-24pt}
\BEA
\NN   \B R_{ij}^+&=&d\omegap_{ij}+\omegap_{ik}\w\omegap^k{}_j
                    +\omegap_{0i}\w\omegap_{0j}\\
                 &=&d\omegap_{ij}+2\omegap_{ik}\w\omegap^k{}_j
\EEA
one defines a connection form on $\S$ as follows
\BE
\label{44}
     A_{ij}:=2i_t^*\omegap_{ij}=\omega_{ij}+i\epsilon_{ij}{}^k\omega_{0k}
\EE
-- here $\omega_{ij}$ and $\omega_{0i}$ denote the
Levi-Civit…-connection and the extrinsic curvature on $\S$ as usual --
because for the curvature $F$ defined by $A$ holds
\BE
      F_{ij}=2i_t^*\Big(d\omegap_{ij}+2\omegap_{ik}\w\B \omegap^k{}_j\Big)
              =2i_t^*\B R_{ij}^+.
\EE
Using the definition one obtains
\BEA
  i_t^*L_{\P/\P\B t}\omegap_{jk}&=&{\TS\F{1}{2}}\dot{A}_{jk}\\
  i_t^*\,^\omegap\!\!Di_{\B N \B e_0}\omegap^{jk}
     &=&\,^A\!Di_t^*i_{\B N\B e_0}\omegap^{jk}
    =:{\TS\F{1}{2}}\,^A\!D Z^{jk}\\
\label{31}
    Z^{jk}=-Z^{kj}&:=&2Ni_t^*i_0\omegap^{jk}
       =a_{jk}+i\epsilon_{jk}{}^lb_l.
\EEA
As usual we pass over to the spatial Lagrange form:
\BEA
\NN   \C L&=& i_t^*i_{\P/\P\B t}\B{\C L}\\
       &=&{\TS\F{1}{2}}\G[i\dot{A}_{jk}\w e^{jk}-iL_\V NA_{jk}\w e^{jk}
         -i\,^A\!DZ_{jk}\w e^{jk}+NF_{ij}\w*e^{jk}\D]
\EEA
At this point one could recognize ${\TS\F{1}{2}}i\epsilon_i{}^{jk}\dot{A}$ as
the canonical momentum of $-{\TS\F{1}{2}}\epsilon^i{}_{jk}e^{jk}$. But this
would be obvious only in a Palatini-like theory where the connection is not
fixed to be the selfdual projection of  the Levi-Civit…-connection. On the
other hand usually one does not vary w.r.t. $A$ in order to obtain the
Levi-Civit…-connection. We are more careful and keep in mind that the
definition of $A$ implies a dependance on the extrinsic curvature, which
is closely linked to the time derivative of the triads. Thus the connection
form $A$ itself depends on $\dot{e}$, so every term of the sum in the
Lagrange form contributes to the calculation of the momentum. To determine
the momentum we write the Lagrangian in the following way
\BEA
\NN
  L(e,\. e, N, \V N)&=&\F{d}{dt}\I\F{i}{2}A_{jk}\w e^{jk}
         -\F{i}{2}\I L_\V N(A_{jk}\w e^{jk})
       -\F{i}{2}\I d(Z_{jk}e^{jk})\\
\label{43}
  &&\hspace{-2.3cm}+\I\G[i\dot{e}^j\w A_{jk}\w e^k-iL_\V Ne^j\w A_{jk}\w e^k
      +iZ_{jk}\,^A\!D e^j\w e^k+\F{N}{2}F_{ij}\w*e^{ij}\D].\quad~~
\EEA
We ignore the boundary terms and notice that one of the terms vanishs in the
\ADM  like Ashtekar theory:
\BEA
\NN  Z_{jk}De^j\w e^k&=&Z_{jk}\,^\omega\!D e^j\w e^k
            +iZ_{jk}\epsilon^{mj}{}_l\omega_{0m}e^l\w e^k\\
\NN  &=&Z_{jk}\omega_{0m}\w*e^{mj}\w e^k\\
     &=&Z_{jk}(\GA \omega_0{}^k|e^j\DA-\eta^{jk}\GA
     \omega_{0m}|e^m\DA)\eta=0
\EEA
Considering this result we are not surprised since an explicit dependance on
$Z_{ij}$ would mean an explicit dependance on the rotational
parameter $a_{ij}$ and $b_i$ defined in (\ref{13}) and (\ref{19}), which
does not appear in the \ADM  Lagrangian. But there is a dependance on
${\dot{e}^i}_{\SSS A}$ (\ref{47}) contrary to the \ADM  theory in the first
term:
\BEA
\NN  i\dot{e}^j\w A_{jk}\w e^k&=&i\. e^j{}_A\w A_{jk}\w e^k
       + i\.e^j{}_S\w A_{jk}\w e^k\\
\NN   &=& i\.e^j{}_A\w \omega_{jk}\w e^k
       + i\.e^j{}_S\w A_{jk}(e^i,\.e^i{}_S)\w e^k\\
      &=& -i\.e^j{}_A\w de_j+i\.e^j{}_S\w A_{jk}(e^i,\.e^i_S)\w e^k
      \qquad\mbox{and}\\
      A_{jk}&=&\omega_{jk}(e)
        +i\epsilon_{jkl}{\TS\F{1}{N}}(L_\V Ne^l{}_S-\.e^l{}_S)
\EEA
So the Lagrangian depends linearly on $\. e^j{}_A$, but is quadratic in
$\. e^j{}_S$. This means a second class constraint for $\. e^j{}_A$, so that
one can disregard this degree of freedom as one knows from the \ADM  theory.
We can guess at this point that the canonical momentum which we derive now
has an additional contribution $-ide^j$ and hence $\GA p_i|*\!e_j\DA$ is no
longer symmetric. We now derive the momentum-2-form.
\BE
   L=\I i\. e^j\w A_{jk}\w e^k-iL_\V Ne^j\w A_{jk}\w e^k
       +\F{N}{2}F_{ij}\w*e^{ij}
\EE
\BD
   \delta_{\. e^i}\I i\. e^j\w A_{jk}\w e^k=i\I \delta\. e^i\w A_{ik}\w e^k+
       i\I \. e^j\w\delta_{\.e^i}A_{jk}(e,\. e)\w e^k
\ED
\BEAS
   i\I\alpha^j\w \delta_{\. e^i}A_{jk}(e,\. e)\w e^k
    &=&-{\TS\F{1}{N}}\I\delta_{\. e^i}\G({\. e^l}_S\w \alpha^j\w *e_{lj}\D)\\
    &=&-{\TS\F{1}{N}}\I\delta_{\. e^i}\G(\. e^l\w {\alpha^j}_S\w*e_{lj}\D)\\
    &=&-{\TS\F{1}{N}}\I\delta\. e^i\w{\alpha^j}_S\w*e_{ij}\qquad,
\EEAS
where $\alpha^i\in\Omega^1(\S)$ and ${\alpha^i}_S={\TS\F{1}{2}}(\alpha^i+
\GA \alpha^j|e^i\DA e_i)$ is defined as usual. Hence we obtain
\BEAS
  \F{\delta}{\delta\. e^i}\I i\.e^j\w A_{jk}\w e^k
     &=& iA_{ij}\w e^j-\F{1}{N}\. e^j{}_S\w*e_{ij}\\
  \F{\delta}{\delta\. e^i}\I iL_\V Ne^j\w A_{jk}\w e^k
     &=& \F{1}{N}L_\V N e^j{}_S\w*e_{ij}\\
  \F{\delta}{\delta\. e^i}\I \F{N}{2}F_{jk}\w*e^{jk}
     &=& \F{\delta}{\delta e^i}\I\F{N}{2}(i_t^*\B R_{jk}
       +i\epsilon_{jk}{}^li_t^*\B R_{0l})\w*e^{jk}\\
     &=&\F{\delta}{\delta\. e^i}\I\bigg[\F{N}{2}
      (R_{jk}+\omega_{0j}\w\omega_{0k})\w*e^{jk}+iNi_t^*
      \underbrace{(\B R_{0l}\w \B e^l)}_{0}\bigg]\\
     &=&-\omega_{0j}\w*e_i{}^j
\EEAS
\BE
\label{14}
 p_i=\F{\delta L}{\delta\. e^i}=iA_{jk}\w e^k
    -\F{1}{N}(\. e^j{}_S-L_\V Ne^j{}_S+N\omega_0{}^j)\w*e_{ij}=
    iA_{jk}\w e^k\qquad.
\EE
We compare this with the momentum 2-form of the \ADM  theory
\BE
\label{36}
  p_i=iA_{ij}\w e^j=i\omega_{ij}\w e^j-\epsilon^k{}_{ij}\omega_{0k}\w e^j
     =-ide_i+p_i^{\mbox{\tiny ADM }}
\EE
and notice that the new momentum does not transform homogeneous under
rotations. As in the \ADM  theory the equation (\ref{14}) can only be solved
on a submanifold of the phasespace on which the following condition holds:
\BEA
\NN  &0~=~iA_{ij}\w e^j-p_i~=~-ide_i+\omega_{0l}\w*e_i{}^l-p_i&\\
\NN  &\Lra\qquad (p_i+ide_i)\w e_j-(p_j+ide_j)\w e_i&\\
     &~=~p_i\w e_j-p_j\w e_i+ id(e_{ij})~=~0&
\EEA
Our Hamiltonian is only determined up to this constraint, so we obtain:
\BEA
\NN   H(e^i,p_i,N,\V N,Z)
         &=&\I\. e^i\w p_i-L-\I Z_{ij}(p^i\w e^j+i de^i\w e^j)\\
\label{21}
       &=&\I\G[p_i\w L_\V N e^i-Z_{ij}(p^i+ i de^i)\w e^j
          -{\TS\F{N}{2}}F_{ij}(p)\w *e^{ij}\D]
\EEA
We call the arbitrary integration function of the rotational constraint $Z$,
because if we derive the equations of motion and want to compare the
Hamiltonian equation with the geometrical equation obtained by
$i_t^*L_{\P/\P\B t}$ we have again to identify the integration parameter
with a space-time quantity by equation (\ref{31}).
We will see in a moment that this Hamiltonian is not suitable for further
consideration, since the substitution of the curvature form $F_{ij}$ by
momentum terms ends up in a lengthy expression. The substitution of $A$ by
$p$ on the constraint manifold yields
\BE
  A_{ij}=-i(\GA p_k| e_{ij}\DA e^k-{\TS\F{1}{2}}p*\!e_{ij})
     \qquad\qquad p:=\GA p_i|*\!e^i\DA.
\EE
For convenience we define also
\BE
   A_i:={\TS\F{i}{2}}\epsilon_i{}^{jk}A_{jk}
   \qquad A_{jk}=-i\epsilon_{ij}{}^k A_k
\EE
and analogously $F_i$ and $Z_i$. Then the relation between $A_i$ and $p_i$
reads
\BE
\label{15}
  A_i=\GA p_k|*\!e_i\DA e^k-{\TS\F{1}{2}}pe_i\qquad\Llra\qquad
    p_i=A_k\w *e^k{}_i
\EE
and finally we obtain for the last term of the Hamiltonian
\BEA
\NN  -{\TS\F{N}{2}}F_{ij}\w*e^{ij}&=&iNF_k\w e^k\qquad\qquad\mbox{\small
                modulo exact forms}\\
\NN  &=&      {\TS\F{N}{2}}\G(\GA p_i|*\!e^j\DA\GA
           p_j|*\!e^i\DA-{\TS\F{1}{2}}p^2
        +2i\GA p_i|*\!e_j\DA\GA de^j|*\!e^i\DA
      -ip\GA de_i|*e^i\DA\D)\eta\\
\label{46}
      &&-i\GA dN|e_i\DA p_j\w*e^{ij}.
\EEA
This is quite an awkward expression for deriving the Hamiltonian equations,
but having in mind that
\BE
   {\TS\F{N}{2}}R\eta={\TS\F{N}{2}}\G[-d(e^i\w*de_i)
      -{\TS\F{1}{2}}(de_i\w e^j)\w*(de_j\w e^i)
      +{\TS\F{1}{4}}(de_i\w e^i)\w*(de_j\w e^j)\D]
\EE
one can easily see that one could have obtained this Hamiltonian by a
canonical transformation $(e^i,p_i^{\mbox{\tiny ADM }})\lmt
(e^i,p_i^{\mbox{\tiny ADM }}-ide_i)=:(e^i,p_i)$ of the \ADM  Hamiltonian
(\ref{20}). So it is not surprising that the Hamiltonian equations
one could derive from (\ref{21}) are really equivalent to the \ADM  equations
of motion, but they contain even more terms than in the \ADM  case. One notices
that the canonical transformation is only possible for spatial dimension 3,
since otherwise the 2-form $de_i$ can not be added to the $n-1$-form $p_i$.
The gauge group $SO(3)$ acts on the configuration space which is
invariant under the canonical transformation. But wheras the action of the
group to the momentum can in the \ADM case be obtained by the obvious
lift to the phase space, one has to transform also the group action to the
new phase space in order to obtain the correct action of the group to the
new momentum
\BEA
\NN
  &e^i\lmt S^i{}_j e^j\qquad
   p^{\mbox{\tiny ADM}}_i\lmt {S^{-1}}^j{}_ip^{\mbox{\tiny
   ADM}}_j=S_i{}^jp^{\SSS
   ADM}_j&\\
\label{34}
      &\qquad p_i\lmt S_i{}^jp_j-idS_i{}^j\w e_j\qquad S\in \S\times SO_C(3)&
\EEA
In order to simplify the expression one considers again the
Lagrange form and notices:
\BE
    L=\I\G[-{\TS\F{i}{2}}A_{jk}\w (e^{jk})^\cdot
    +{\TS\F{i}{2}}A_{jk}\w L_\V Ne^{jk}+{\TS\F{N}{2}}F_{ij}\w*e^{ij}\D]
\EE
One would like to perform another canonical transformation which turns the
conection form $A$ itself into the momentum. We now define
and would like to determine the canonical conjugate $q^i\in \Omega^2(\S)$
such that
{\arraycolsep=1.4mm
\BEAS
  &&\bigg\{\I q^i\w \alpha_i,\I A_j\w\beta^j\bigg\}\\
   &=&\I\Bigg[\bigg(\F{\delta}{\delta e^k}\I q^i\w\alpha_i\bigg)\w
       \bigg(\F{\delta}{\delta p_k}\I A_j\w \beta^j\bigg)
      -\bigg(\F{\delta}{\delta p_k}\I q^i\w \alpha_i\bigg)\w
       \bigg(\F{\delta}{\delta e^k}\I A_j\w \beta^j\bigg)\Bigg]\\
      &=&\I \alpha_i\w\beta^i
\EEAS}
where $\alpha_i\in\Omega^1(\S)$ and $\beta^j\in\Omega^2(\S)$. Using
equation (\ref{15}) one can calculate the derivatives of the second term.
\BEAS
  \F{\delta}{\delta p_k}\I A_j\w\beta^j
     &=&e_j\GA \beta^j|*\!e_k\DA-{\TS\F{1}{2}} e^k\GA \beta_j|*\!e^j\DA\\
  \F{\delta}{\delta e^k}\I A_j\w \beta^j
     &=&\beta^j\GA p_k|*\!e_j\DA-\GA\beta^j|*\!e^i\DA i_kp_i\w e_j
     -{\TS\F{1}{2}}\beta_k p
     +{\TS\F{1}{2}}i_kp_j\w e^j\GA \beta_i|*\!e^i\DA
\EEAS
One can now guess $\F{\delta}{\delta p_k}\I q^i\w \alpha_i=0$, because
otherwise one can hardly get rid of the momentum dependence. Consequently
one supposes for the canonical coordinate $q^i=c*\!e^i$ and obtains
\BEAS
    &\DS\F{\delta}{\delta e^k}\I q^i\w \alpha_i
      =c\epsilon^i{}_{kl}e^l\w\alpha_i=c*e^i{}_k\w \alpha_i\qquad\mbox{and}&\\
    &\DS c*e^i{}_k\w\alpha_i\w\G(e_j\GA \beta^j|*e^k\DA
        -{\TS\F{1}{2}}e^k\GA\beta_j|*e^j\DA\D)=-c\alpha_i\w\beta^i&
\EEAS
so our new configuration variable is
\vspace{-12pt}
\BE
   \hspace{35mm}q^i=-*e^i.
\EE
Introducing coordinates $\{x^a\}_{a=1,2,3}$ for a moment we can link these
varibles with the densities used in Ashtekar's formulation. Since in
integrals one has expressions like
\BD
    \I q^i\w\alpha_i=
      {\TS\F{1}{2}}\I dx^a\w dx^b\w dx^c q^i_{ab}{\alpha_i}_c
    ={\TS\F{1}{2}}\I d^3x\epsilon^{abc}q^i_{ab}{\alpha_i}_c
\ED
we consider
\BE
  {\TS\F{1}{2}}\epsilon^{abc}q^i_{ab}
  =-{\TS\F{1}{4}}\eta^{il}\epsilon^{abc}\epsilon_{jkl}e^{jk}_{ab}
  =-{\TS\F{1}{2}}\eta^{il}\epsilon^{abc}\epsilon_{jkl}e^j_ae^k_b
  =-\eta^{il}e^c_l\det e^i_a =\F{e^c_l}{\det e^a_i}=-\eta^{il}E^c_l
\EE
where $e^i_a=e^i(\P/\P x^a),~e^a_i=dx^a(e_i)$ and thus
$e^i_ae^a_j=\eta^i_j$. So the densities $E^c_l$ are just the coordinate
expressions of the 2-forms $q^i$ -- up to a sign and an $\epsilon$-symbol
usually hidden in the volume $d^3x$. The densities $E^a_i$ determine
directly the coordinates of the dual triads
\BE
\label{38}
    e^a_i=\F{E^a_i}{\det^{1/2}E^a_i}
\EE
and these of course determine the triads itself
\BE
\label{39}
    e^i_a=\F{1}{2\det e^a_i}\epsilon^{ijk}\epsilon_{abc}e^b_je^c_k
         =\F{1}{2\det^{1/2}E^a_i}\epsilon^{ijk}\epsilon_{abc}E^b_jE^c_k
\EE
what insures that the metric can be reconstructed from the new variables.

Substituting $(e^i,p_i)$ in the Hamiltonian by $(q^i,A_i)$ one obtains
\BEA
\NN  H(q^i,A_i)&=&\I A_k\w*e^k{}_i\w L_\V Ne^i
       -Z_{ij}(A_k\w*e^k{}_i+ide^i)\w e^j
       -{\TS\F{N}{2}}\epsilon^{ij}{}_kF_{ij}\w e^k\\
     &=&\I A_k\w L_\V Nq^k+Z_kDq^k-iNF_k(A)\w *q^k
\EEA
where
\vspace{-24pt}
\BEA
\NN   Dq^k&=&dq^k+A^k{}_j\w q^j=dq^k-i\epsilon^k{}_{jl}A^l\w q^j\\
      D\V q&=&d\V q+i\V A\wk \V q
\EEA
and
\vspace{-24pt}
\BEA
\NN    F_k&=&dA_k+{\TS\F{i}{2}}\epsilon_k{}^{ij}A_i\w A_j \\
       \V F&=& d\V A+{\TS\F{i}{2}}\V A\wk \V A
\EEA
Using the vector notation we obtain what we will call the Ashtekar
Hamiltonian:
\BE
  H(\V q,\V A)=\I\G[\V A\wm L_\V N\V q-\V Z\wm D\V q-iN \V F\wm *\V q\D]
\EE
One easily recognizes the three terms as diffeomorphism, rotational and
Hamiltonian part. The derivation of the equations of motion is simple up to
a single point: We have to derive a term $\I\V \alpha\wm*\V
q=\I\alpha_i\w*q^i$ with respect to the 2-form $q^j$. Reexpressed by the
triads the problem reads:
\BD
  \F{\delta}{\delta*\!e^i}\I e^m\w\alpha_m\qquad \alpha_i\in\Omega^2(\S)
\ED
Since there are as many independent dual triads as triads itself for any
space dimension and one can reexpress the metric in terms of coordinates of
the dual forms $*e^i$ as well as in terms of the triads the problem has a
simple solution which we described in (\ref{16}). Then we find the following
equations of motion:
\BEA
   \.{\V q}=\F{\delta H}{\delta\V A}
     &=&L_\V N\V q-i\V Z\times\V q-iD(N*\!\V q)\\
\label{22}
   \.{\V A}=-\F{\delta H}{\delta\V q}
     &=&L_\V N\V A+D\V Z-N\R-{\TS\F{N}{4}}F*\!\V q
\EEA
Here $i_jF^j{}_i=:(\R)_i$ is the Ricci-form and $F=\GA F_{ij}|e^{ij}\DA$ is
the Ricci-scalar of $F$. These equations of motion are the complex
extensions of the \ADM  equations as can be checked using the following
relations
\BEA
\label{25}
    A_i&=&{\TS\F{i}{2}}\epsilon_i{}^{jk}A_{jk}=
       {\TS\F{i}{2}}\epsilon_i{}^{jk}\omega_{jk}-\omega_{0i}\\
\label{17}
     Z_i&=&{\TS\F{i}{2}}\epsilon_i{}^{jk}Z_{jk}
       ={\TS\F{i}{2}}\epsilon_i{}^{jk}a_{jk}-b_i
       ={\TS\F{i}{2}}\epsilon_i{}^{jk}a_{jk}+\GA dN|e_i\DA
\EEA
In order to check the equations one needs the equation of motion for
$\omega_{0i}$ which can either be obtained from the \ADM  equations of
motion or by use of the equation of motion for the extrinsic curvature
$K_{ab}$, where the indices are w.r.t. a fixed (coordinate) basis
$\{dx^a\}_{a=1,2,3}$\cite{schirm}:
\BEAS
  \. K_{ab}&=&L_\V NK_{ab}+2NK_{ac}K^c{}_b
    -NKK_{ab}\\
    &&-N\mbox{Ric}R_{ab}+\nabla\!_a\nabla\!_bN
      +{\TS\F{N}{4}}(R-K_{cd}K^{cd}+K^2)h_{ab}
\EEAS
\BD
  h_{ab}=\eta_{ij}e^i(\P_a)e^j(\P_b)
    \qquad\qquad\omega_{0i}=-K_{ab}dx^be^a{}_i
    \qquad\qquad e^a{}_i=dx^a(e_i)
\ED
Using
\vspace{-24pt}
\BEAS
      \. e^a{}_i&=&dx^a(\. e_i)=-dx^a(e_j)\. e^j(e_i)\\
      \. e^j&=&L_\V Ne^j-a^j{}_k e^k-N\omega_0{}^j
\EEAS
one finally obtains
\BEA
\label{23}
\NN \. \omega_{0i}&=&-(K_{ab}e^a{}_i)^\cdot dx^b\\
\NN   &=&L_\V N\omega_{0i}-D\GA dN|e_i\DA +a^k{}_i\omega_{0k}+\\
      &&N(\mbox{Ric}R_{ij}-K_{ik}K^k{}_j+KK_{ij})e^j
      -{\TS\F{N}{4}}(R-K_{kl}K^{kl}+K^2)\eta_{ij}e^j.
\EEA
It is nearly obvious that this equation is the real part of equation
(\ref{22}), if one regards all differential geometric objects as real,
i.e. $e^i,\V N,N, a^i{}_j,\omega_{0i}$. Regarding the equations of
motion and having in mind the relation $\GA dN|e_i\DA+b_i=0$ and (\ref{17})
one can easily understand why in the equation of motion for $\. q$ there
appears a derivative term of the lapse contrary to the \ADM  theory and why
there is no derivative term for the lapse in the equation for $A$. If one
decomposed $Z$ into real and imaginary part another derivative term for the
lapse would appear. We notice a difference to the usual Ashtekar
formulation. There is no need to densitize our lapse function. If
we had done so, the last term of equation (\ref{22}) containing the
Ricci-Scalar of $F$ would vanish. This is not dramatic since we will see in
a moment that the Hamiltonian constraint in the Ashtekar formulation is just
$F=0$ and since our constraint algebra will prove first class we will never
leave the constraint surface. The Ricci-scalar-term corresponds
to the last term in equation (\ref{23}) which we recognize as the the
the Einstein-tensor $G(n,n)$ or the Hamiltonian constraint of the \ADM
formulation here expressed by the extrinsic curvature instead of the
momentum. But since one does usually not omit the terms with $*e_i$ in the
\ADM  equation (\ref{24}) whose sum vanishs on the constraint surface
and in order to perform a correct constraint analysis, we keep the
Ricci-scalar term here. Finally we note that of course $A$ does not
transform homogeneous under rotations as well and that the rotational part
in the equation of motion for $A$ could have been derived easily from
the canonical transformation of the $SO(3)$-action, using equations
(\ref{15}) and (\ref{34}).

Finally we will perform the full constraint analysis. We have already
encountered the first primary constraint, the rotational constraint. The
secondary constraints associated with lapse and shift read -- as usual we
disregard all boundary terms:
\BEA
\NN   {C_D}_i&=&\{H,p_{N^i}\}=\F{\delta}{\delta N^i}\I A_j\w L_\V Nq^j\\
\NN          &=&A_j\w i_idq^j+dA_j\w i_iq^j\\
             &=&i_iA_jDq^j+F_j\w i_iq^j\\
       C_H&=&\{H,p_N\}=-i\V F\wm *\V q=-{\TS\F{1}{2}}F\eta
\EEA
Using the identity $i_iq^j=-\epsilon_{ki}{}^j*\!q^k$ and the relation:
\BEA
\NN (\V F\wk*\V q)_i&=&\epsilon_{ijk}\GA F^j| e^k\DA\eta
        =-{\TS\F{i}{4}}\epsilon_{ij}{}^k\epsilon^{jlm}\epsilon_{krs}
          \GA F_{lm}|e^{rs}\DA \eta\\
\label{18}
       &=&i(\R \wk\V q)_i
\EEA
we can write the diffeomorphism constraint as follows
\BE
      {C_D}_i=i_iA_j Dq^j-i(\R \wk\V q)_i
\EE
and the integrated version then reads
\BE
\label{40}
  H_D(\V N)=\I \V A\wm L_\V N\V q=\I\G[\V A(\V N) D\V q-i\V N(\R \wk \V q)\D]
\EE
We reformulate also the rotational constraint in our variables
\BE
    {C_R}^i=Dq^i
\EE
and notice that the differential form of the diffeomorphism constraint can
be simplified using the rotational constraint can be simplified using the
rotational constraint
\BE
   {\T C}_{{\SSS D}\,i}=-i_i\V F\wm\V q=-i(\R\wk\V q)_i
\EE
Usually this version of the diffeomorphism constraint is used in the
Ashtekar formulation, its integrated form reads
\BE
   \T H_D(\V N)=-\I i_\V N\V F\wm \V q=-i\I \V N(\R \wk\V q),
\EE
but we prefer $H_D=\I \V A\wm L_\V N \V q$, because it is just the momentum
mapping of action of the diffeomorphism group on the phase space. Thus it is
not even necessary to calculate the Poisson-bracket between two
diffeomorphism constraints in this form. That the differential part of the
diffeomorphism constraint has a rotational part is not a special feature
of the Ashtekar formalism as we have seen in (\ref{30}). We can write the
Hamiltonian as a sum of integrated constraints, neglecting all boundary
terms:
\BE
   \BA{rcccccc}
     H&=&\DS\I N^i{C_D}_i&+&\DS\I Z_i{C_R}^i&+&\DS\I NC_H\\
     &=&H_D(\V N)&+&H_R(\V Z)&+&H_H(N)
   \EA
\EE
We determine again the constraint algebra, proving that the constraints
are first class:
\BE
  \BA{rcl}
   \{H_D(\V N),H_D(\V M)\}&=&H_D([\V N,\V M])\\
   \{H_D(\V N),H_R(\V Z)\}&=&H_R(L_\V N \V Z)\\
   \{H_D(\V N),H_H(N)\}&=&H_H(L_\V NN)\\
   \{H_R(\V Z_1),H_R(\V Z_2)\}&=&-iH_R(\V Z_1\times \V Z_2)\\
   \{H_R(\V Z),H_H(N)\}&=&0\\
   \{H_H(N),H_H(M)\}&=&H_D(NdM^\sharp-MdN^\sharp)
      -H_R(\GA \V A|NdM-MdN\DA)
  \EA
\EE
The first and the third equation are proved as in the previous sections.
\BEAS
  \{H_D(\V N),H_R(\V Z)\}
    &=&\I\G[ iL_\V N\V A\wm (\V Z\times\V q)+L_\V N\V q\wm D\V Z\D]\\
    &=&\I\G[ iL_\V N\V Z (\V A\times \V q)-L_\V N\V Z\wm d\V q\D]\\
    &=&+\I L_\V N\V Z\wm D\V q=H_R(L_\V N\V Z)\\
  \{H_R(\V Z_1),H_R(\V Z_2)\}
    &=&\I\G[i D\V Z_1\wm(\V Z_2\times\V q)-iD\V Z_2\wm(\V Z_1\times\V q)\D]\\
    &=&-i\I\V (\V Z_1\times\V Z_2)\wm D\V q=-iH_R(\V Z_1\times\V Z_2)\\
  \{H_R(Z),H_H(N)\}
    &=&\I\G[i D\V Z\wm D(N*\!\V q)
      +iN(\V Z\times\V q)(\R +{\TS\F{1}{4}}F*\!\V q)\D]\\
    &=&\IN\G[(\V F\times\V Z)\wm *\V q-i(\R\wk\V q)\cdot\V Z\D]
       \stackrel{(\ref{18})}{=}0\\
  \{H_H(N),H_H(M)\}
    &=&\I\bigg[-iN(\R+{\TS\F{1}{4}}F*\!\V q)\wm D(M*\!\V q)\\
          &&~~~~~~+iM(\R +{\TS\F{1}{4}}F*\!\V q)\wm D(N*\!\V q)\bigg]\\
    &=&\I i(\R\wm*\V q)\w(NdM-MdN)\\
    &=&\I i\R\wk\V q\cdot\GA NdM-MdN|*\!\V q\DA\\
    &\stackrel{(\ref{40})}{=}&
        H_D(NdM^\sharp-MdN^\sharp)-\I\GA \V A|NdM-MdN\DA \wm D\V q\\
    &=&H_D(NdM^\sharp-MdN^\sharp)-H_R(\GA \V A| NdM-MdN\DA)
\EEAS
So these variables permit a simple derivation of the constraint algebra.
At this point we have managed to give an "\ADM  like" Ashtekar formalism
where only the metric was regarded as dynamical variable whereas the
connection was fixed as the Levi-Civit…-connection or its selfdual
projection. In a Palatini-like Ashtekar theory, where also the connection
is regarded as a variable, the solution of the constraints should lead to
equation (\ref{25}) by which the arbitrary connection form is linked to the
Levi-Civit…-connection of the triads and the extrinsic curvature. Our
analysis also gives a simple  interpretation of the reality constraints. We
know that the Ashtekar Hamiltonian is just a canonical transformation of the
complexified \ADM Hamiltonian, so we consider first the complexified \ADM
theory. Since the differential equation is real for a real rotational term
it is clear that a real initial condition has a real time development, so a
pair of real triads and momentum forms $(e^i,p^{\mbox{\tiny ADM}}_i)$ as
initial condition will produce a real development of the metric, even in the
case that the rotational parameter is not real. Let us now suppose that the
spatial metric represented by triads $\eta_{ij}e^i\otimes e^j$ is real for a
certain time. Then there is a SO$_C$(3)-valued function $S$ on
$\setR\times\S$ such that $\T e^i:= S^i{}_je^j$ are real triads, having real
components with respect to a coordinate basis. One can easily see that $(\T
e^i,\T p^{\mbox{\tiny ADM}}_i), ~\T p^{\mbox{\tiny
ADM}}_i:=S_i{}^jp^{\mbox{\tiny ADM}}_j$ satisfy the equations of motion
(\ref{32},\ref{24}) for another rotational term. Considering the split
equation for the triads (\ref{33}) one notices that provided lapse and shift
are real the extrinsic curvature $\T \omega_{0i}$ is real and consequently
the momentum form $\T p^{\mbox{\tiny ADM}}_i$ is real. Thus the most general
initial condition which leads to a real metric is a pair of real forms
$(e^i,p^{\SSS ADM}_i)$ modulo of course a possible complex rotation. One
could write the condition in the form
\BEA
            \eta_{ij}e^i\otimes e^j&=&h~=~\mbox{real}\\
\label{37}
   \eta^{k(i}e_i\otimes e_j\otimes p^{\mbox{\tiny ADM}}_k\w e^{j)}
   &\stackrel{(\ref{41})}{=}&\pi=\mbox{real},
\EEA
but one could as well simplify the second condition by
\BE
            p_i\otimes e^i=\mbox{real}
\EE
and replace the first one by
\BE
            \eta^{ij}p^{\mbox{\tiny ADM}}_i\otimes p^{\mbox{\tiny ADM}}_j
            =\mbox{real}
\EE
Now we have only to transform this condition from the complexified \ADM
theory to the Ashtekar variables. Formally one can restate the reality
conditions as $(q^i,A_i-{\TS\F{i}{2}}\epsilon_i^{jk}\omega_{jk}(q)
=-\omega_{0i}$ is real modulo a complex rotation or
\BEA
     \eta_{ij}q^i\otimes q^j&=&\mbox{real}\\
        q^i\otimes \omega_{0i}(A,q)&=&\mbox{real},
\EEA
but it is quite difficult to obtain $\omega_{ij}$ from $q^i$, so the second
condition is difficult to check given the coordinates of $q^i$ and $A_i$.
The first condition is obviously equivalent to the condition that the
spatial metric $h$ is real, so one can reexpress equation (\ref{37})
in terms of Ashtekar variables using
\BE
   p^{\mbox{\tiny ADM}}_i=i\eta_{ij}D e^j=-i \eta_{ij}D *\!q^j,
\EE
which is a simple consequence of equations (\ref{14}) and (\ref{36}).
If one introduces coordinates and uses the expressions (\ref{38}) and
(\ref{39}) for the coordinates of triads and dual triads in terms of the
densities one finds
\BEAS
   \pi^{ab}{}_{123}&=&\epsilon^{cde}(e_i\otimes e_j\otimes p^{(i}\w e^{j)})
                (dx^a,dx^b,\P_c,\P_d,\P_e)\\
     &=&-i\F{1}{\det E^a_i}(\V E^f\times D_f\V E^{(a})
     \cdot \V E^{b)}~=~\mbox{real}
\EEAS
and this is up to the factor $-1/\det E^i_a$, which is clearly real,
the coordinate expression for the second reality constraint, which is
usually derived by requiring that the Hamiltonian flow leaves the metric
real {\cite{Guilini}}. So the reality constraints given in the Ashtekar
theory just state that the reconstructed \ADM metric and momentum is real.

Our analysis has shown that it might be useful to perform calculations in
canonical gravity as far as possible without reference to a fixed reference
frame. We have seen in which way the coordinate free version of Ashtekar's
formulation is related to the complexified \ADM theory and how the
rotational parameters of the complexified theory are linked to the
real theory. One might ask if  triads have a real physical significance or
if they are only useful tools for deriving equations for the  metric. In
Minkowski space one usually derives the energy momentum tensor by
considering the variation of the Lagrangian as a consequence of a
translation in space-time which could be written as a derivation with
respect to the 1-form $dx^\mu$. The obvious generalisation for a curved
spacetime is the variation w.r.t. the four orthogonal 1-forms $e^\mu$ what
really yields the energy-momentum-tensor. This could be seen as an argument
that the tetrades really are physically significant.

{\footnotesize{\bf Acknowledgment:} I would like to thank P.
Gl\"o\ss ner for help and motivation.}



\begin{thebibliography}{xxx}
\bibitem{ThII}W. Thirring {\it Lehrbuch der mathematischen Physik Bd. 2}
           Springer Wien, New York 1990
\bibitem{Palatini}A. Palatini, Rend. Circ. Math. Palermo 43, 203 (1919)
\bibitem{schirm}J. Schirmer {\it Das Wirkungsprinzip in der allgemeinen
           Relativit\"atstheorie} diploma thesis Freiburg 1994
\bibitem{Hehl}F.
           Hehl, P. von der Heyde, G. D. Kerlick, J. M. Nester,
           {\it General Relativity with Spin and Torsion: Foundations
           and Prospects} Rev. Mod. Phys., Vol 48, 3, 393-416,
           1976
\bibitem{Dirac}P. A. M. Dirac {\it Lectures of Quantum Mechanics}
           Yeshiva University Press, New York 1964
\bibitem{Ashb} A. Ashtekar {\it Lectures on Non-Pertubative Canonical
           Gravity,} World Scientific, Singapur 1991
\bibitem{Guilini} D. Guilini {\it Ashtekar Variables in Classical General
           Relativity} in {\it Canonical Gravity: From Classical to Quantum}
           Ed. J. Ehlers, H. Friedrich, Springer, Berlin Heidelberg 1994
\bibitem{A&M} R. Abraham, J. E. Marsden {\it Foundations of Mechanics,}
           Benjamin/Cummings, Reading Massachusetts 1978
\bibitem{5}R. Arnowitt, S. Deser, C. W. Misner {\it The Dynamics of General
           Relativity}  in: {\it Gravitation An Introduction to Current
           Research} ed. L. Witten, Wiley, New York 1962
\end{thebibliography}
\enddocument